\documentclass[12pt,a4paper]{article}
\usepackage[margin=1in]{geometry}
\usepackage{setspace}
\usepackage{authblk}

\usepackage{microtype} 
\usepackage{amsthm}
\usepackage{amssymb,amsmath}
\usepackage{natbib}
\usepackage[colorlinks,citecolor=blue,urlcolor=blue,filecolor=blue,backref=page]{hyperref}
\usepackage{graphicx}
\usepackage{algorithm}
\usepackage{algorithmic}
\usepackage{subfigure} 
\usepackage{comment}

\newcommand{\Expect}[1]{\mathbb{E} \left[{#1}\right]}
\newcommand{\Expects}[2]{\mathbb{E}_{{#1}} \left[{#2}\right]}

\newcommand{\Vars}[2]{\mbox{Var}_{{#1}} \left[{#2}\right]}

\newcommand{\md}{\mbox{d}}
\newcommand{\GP}{\mathcal{L}}
\newcommand{\gp}{\ell}

\newcommand{\pihat}{\hat{\pi}}

\begin{document}

\title{Merging MCMC Subposteriors through Gaussian-Process Approximations}

\author{Christopher Nemeth and Chris Sherlock}
\affil{Department of Mathematics and Statistics, Lancaster University, U.K.}

\maketitle
\begin{abstract}
Markov chain Monte Carlo (MCMC) algorithms
have become powerful tools for Bayesian inference. However, they do
not scale well to large-data problems. 
 Divide-and-conquer strategies, which split the data into batches and,
for each batch, run independent MCMC algorithms targeting the
corresponding subposterior, can spread the computational burden across
a number of separate computer cores. The challenge with such strategies is in
recombining the subposteriors to approximate the full posterior. By
creating a Gaussian-process approximation for each log-subposterior density we
create a tractable approximation for the full posterior. This
approximation is exploited through three methodologies: firstly
 a Hamiltonian Monte Carlo algorithm targeting the
expectation of the posterior density provides a sample from an
approximation to the posterior; secondly, evaluating the true posterior at the sampled points leads to an importance sampler that,
asymptotically, targets the true posterior expectations; finally, an
alternative importance sampler uses the full Gaussian-process distribution of
the approximation to the log-posterior density to re-weight any initial
sample and provide both an estimate of the posterior expectation and a measure of the
uncertainty in it.
\end{abstract}


\section{Introduction}
\label{sec:introduction}

Markov chain Monte Carlo (MCMC) algorithms are popular tools for
sampling from Bayesian posterior distributions in order to estimate
posterior expectations. They benefit from
theoretical guarantees of asymptotic convergence of the estimators as
the number of MCMC samples grows. However, whilst asymptotically exact,
they can be computationally expensive when applied to datasets with a
large number of observations $n$. Indeed, the cost of generating one
sample from the MCMC algorithm is at best $O(n)$ as the posterior distribution
of the model parameters, conditional on the entire data set, must be
evaluated at each MCMC iteration. For very large $n$, therefore, MCMC
algorithms can become computationally impractical.

Research in the area of MCMC for big data can be broadly split into
two streams: those which utilise one core of the central processing
unit (CPU) and those that distribute the work load across multiple
cores, or machines. For the single processor case, 
the computational cost of running MCMC on the full data set may be
reduced by using a random subsample of the data at each iteration
\citep{Quiroz2014,Maclaurin2014,Bardenet2014}; however, the mixing of
the MCMC chain can suffer as a result. 
Alternatively, the Metropolis-Hastings acceptance step can be avoided
completely by using a stochastic gradient algorithm
\citep{Welling2011,Chen2014}, where subsamples of the data are used to
calculate unbiased estimates of the gradient of the log-posterior. Consistent estimates of
posterior expectations are obtained as the gradient step-sizes
decrease to zero \citep{Teh2014}. While popular, subsampling methods
do have the drawback that the data must be independent and the whole data set must be readily available at all
times, and therefore data cannot be stored across multiple machines.

Modern computer architectures readily utilise multiple cores of the
CPU for computation, but MCMC algorithms are inherently serial in
implementation. Parallel MCMC, where multiple MCMC chains, each targeting
the full posterior, are run on
separate cores, or machines, can be easily executed
\citep{Wilkinson2005}, however, this does not address the big-data
problem as each machine still needs to store and evaluate the whole
data set. In order to generate a significant computational speed-up,
the data set must be partitioned into disjoint batches, where
independent MCMC algorithms are executed on separate batches on
independent processors \citep{Huang2005}. Using only a subset of the
entire data means that the MCMC algorithm is targeting a different posterior distribution conditional on only a subset of the data, herein referred to as a \textit{subposterior}. This type of
parallelisation is highly efficient as there is no communication
between the parallel MCMC chains. The main challenge is to then
reintegrate the samples from the separate MCMC chains to approximate
the full posterior distribution. \citet{Scott2013} create a Gaussian
approximation for the full posterior by taking weighted averages of
the means and variances of the MCMC samples from each batch; this
procedure is exact when each subposterior is Gaussian, and can work well approximately in non-Gaussian scenarios. \citet{Neiswanger2013} avoid the Gaussian assumption by approximating the subposteriors using kernel density estimation, however, kernel density approximations scale poorly in high dimensions \citep{Liu2007a}. Also, the upper bounds on the mean squared error given in \citet{Neiswanger2013} grow exponentially with the number of batches, which is problematic in big data scenarios where the computational benefit of parallelisation is proportional to the number of available processors.

Previous approaches used to merge the product of subposterior densities have solely relied on the parameter samples outputted from each MCMC algorithm,
but have neglected to utilise the subposterior densities which
are calculated when evaluating the Metropolis-Hastings ratio. 
We place Gaussian-process (GP) priors on the log-density of
each subposterior. The resulting approximation to the log of the full
posterior density is a sum of Gaussian-processes, which is itself a Gaussian-process, assuming that each individual GP has finite variance. Using this formulation we can obtain point estimates of any expectation of interest. The uncertainty in these point estimates is captured by the covariance of the GP approximation to the posterior.

Starting from this Gaussian-process approximation to the full
log-posterior density, we investigate three approaches to
approximating the posterior. Firstly, an efficient Hamiltonian Monte
Carlo (HMC) algorithm \citep{Neal2010} which targets the expectation of the
posterior density (the exponential of the combined GP); samples from this
provide our first means of estimating expectations of
interest. Secondly, the HMC sample values may be sent to each of the cores,
with each core returning the true log-subposterior at each of the
sample points. Combining these \textit{coincident} log-subposterior values
provides the true posterior at the sampled points, which in turn
provides importance weights for the HMC sample, leading to
asymptotically consistent estimates of posterior
expectations. Alternatively, one may wish to avoid the computational expense of running HMC on the expectation of the
exponential of the GP, and of calculating the true sub-posteriors at a sample
of points. We, therefore, also consider an importance proposal based upon any
approximation to the true posterior and obtain repeated samples of importance
weights by repeatedly sampling realisations of the GP approximation to
the log-posterior. This provides both an estimate of any expectation
of interest and a measure of its uncertainty.

This paper is structured as follows. Section
\ref{sec:bayes-infer-with} reviews the divide-and-conquer MCMC approach for
sampling from the posterior, the HMC algorithm and importance
sampling. Section \ref{sec:gaussian-processes} then
outlines the creation of our Gaussian-process approximation for each of the individual
subposteriors, and for combining these. In Section
\ref{sec:three-methods} we detail three methods for approximating
 posterior expectations, each utilising the combined Gaussian-process approximation. 
Section \ref{sec:simulation-study} provides a numerical comparison of our proposed method against four alternative algorithms. We consider five different statistical models leading to a range of posterior distributions. We compare our proposed method against competing algorithms from the literature, and show how our method is particularly successful at approximating non-Gaussian posteriors.

\section{Bayesian inference and MCMC}
\label{sec:bayes-infer-with}

Consider a data set $\mathcal{Y}=\{y_1,y_2,\ldots,y_n\}$ where we assume that the data are conditionally independent with a likelihood $\prod_{i=1}^n p(y_i|\vartheta)$, where $\vartheta \in \Theta \subseteq \mathbb{R}^d$ are model parameters. Assuming a prior $p(\vartheta)$ for the parameters, the posterior distribution for $\vartheta$ given $\mathcal{Y}$ is 
\begin{equation}
  \label{eq:1}
  \pi(\vartheta) = p(\vartheta|\mathcal{Y}) \propto p(\vartheta)\prod_{i=1}^n p(y_{i}|\vartheta).
\end{equation}

Alternatively, the data set $\mathcal{Y}$ can be partitioned into $C$ batches $\{\mathcal{Y}_1,\mathcal{Y}_2,\ldots,\mathcal{Y}_C\}$ where we define a \textit{subposterior} operating on a subset of the data $\mathcal{Y}_c$ as 
\begin{equation}
  \label{eq:3}
    \pi_c(\vartheta) = p(\vartheta|\mathcal{Y}_c) \propto p(\mathcal{Y}_{c}|\vartheta)p(\vartheta)^{1/C},
\end{equation}
where $p(\vartheta)$ is chosen so that $p(\vartheta)^{1/C}$ is proper. The full posterior is given as the product of the subposteriors $\pi(\vartheta) \propto \prod_{c=1}^C \pi_c(\vartheta)$. In this setting we no longer require conditional independence of the data, but rather independence between the batches $\{\mathcal{Y}_c\}_{c=1}^C$, where now the data in each batch can exhibit an arbitrary dependence structure.

Creating an approximation to the posterior, $\pi(\vartheta)$, commences with sampling from
each of the subposteriors $\pi_c(\vartheta)$ independently in parallel,
where, given the independence between data subsets, there is no
communication exchange between the MCMC algorithms operating on the
subposteriors. This type of parallelisation is often referred to as
\textit{embarrassingly parallel} \citep{Neiswanger2013}. The challenge
then lies in \textit{combining} the subposteriors, for which we propose using
Gaussian-process approximations. 

In this paper, we introduce the Hamiltonian Monte Carlo (HMC) algorithm
as one possible MCMC algorithm that can be used to sample from
$\pi_c(\vartheta)$. Moreover, we use HMC in Section
\ref{sec:three-methods} to sample from an approximation to the full posterior, $\pi(\vartheta)$. Other MCMC
algorithms, including the random walk Metropolis \citep{Robert1997},
Metropolis adjusted Langevin algorithm \citep{Roberts1998} and
adaptive versions of these \citep[e.g.][]{Andrieu2008} can also be used. 

\subsection{Hamiltonian Monte Carlo}
\label{sec:background}

We now provide a brief overview of Hamiltonian Monte Carlo and its 
application in this paper; the interested reader is referred to
\citet{Neal2010} for a full and detailed review. The HMC algorithm
considers the sampling problem as the exploration of a physical system
with $-\log \pi(\vartheta)$ corresponding to the potential energy at
the position $\vartheta$. We then introduce artificial momentum
variables $\varphi \in \mathbb{R}^D$, with 
$\varphi \sim \mathcal{N}(0,M)$ being independent of $\vartheta$. Here $M$ is a mass matrix that can be set to the identity matrix when there is no information about the target distribution. This scheme now augments our target distribution so that we are now sampling $(\vartheta,\varphi)$ from their joint distribution
\begin{equation}
  \label{eq:jointDist}
\pi(\vartheta,\varphi) \propto \exp \left( \log \pi(\vartheta) -\frac{1}{2}\varphi^\top M^{-1} \varphi \right),
\end{equation}
the logarithm of which equates to minus the total energy of the system.
Samples from the marginal distribution of interest, $\pi(\vartheta)$, are obtained by discarding the $\varphi$ samples. 

We can sample from the target distribution by simulating $\vartheta$ and $\varphi$ through fictitious time $\tau$ using Hamilton's equations (see \citet{Neal2010} for details)
\begin{equation}
  \label{eq:hamiltonDynamics}
  d\vartheta = M^{-1} \varphi d\tau, \quad\quad   d\varphi = \nabla_\vartheta \log \pi(\vartheta) d\tau.
\end{equation}

The differential equations in \eqref{eq:hamiltonDynamics} are intractable and must be solved numerically. Several numerical integrators are available which preserve the volume and reversibility of the Hamiltonian system \citep{Girolami2011}, the most popular being the \textit{leapfrog}, or Stormer-Verlet integrator. The leapfrog integrator takes $L$ steps, each of size $\epsilon$, on the Hamiltonian dynamics \eqref{eq:hamiltonDynamics}, with one step given as follows:
\begin{eqnarray*}
\varphi_{\tau+\frac{\epsilon}{2}} &=& \varphi_\tau + \frac{\epsilon}{2}\nabla_{\vartheta_\tau} \log \pi(\vartheta_\tau) \\
\vartheta_{\tau+\epsilon} &=& \vartheta_{\tau} + \epsilon M^{-1} \varphi_{\tau+\frac{\epsilon}{2}} \\
\varphi_{\tau+\epsilon} &=& \varphi_{\tau+\frac{\epsilon}{2}} + \frac{\epsilon}{2}\nabla_{\vartheta_\tau+\epsilon} \log \pi(\vartheta_{\tau+\epsilon}) \\  
\end{eqnarray*}
Using a discretisation introduces a small loss or gain in the total
energy, which is corrected through a Metropolis-Hastings accept/reject
step. 
%This reduces the acceptance rate from one, in the ideal analytic
%case, to a rate that is usually higher than the random walk
%Metropolis and results in the acceptance of proposals which are far
%from the current position in the Markov chain. 
The full HMC algorithm is given in Algorithm \ref{alg:hamiltonian} in
Appendix \ref{app.HMC.alg}.

The HMC algorithm has a step-size parameter $\epsilon$ and number of leap frog steps $L$ which need to be tuned. The performance of the algorithm is highly dependent on the tuning of the parameters. One way to tune the algorithm is to optimise the parameters such that the acceptance rate is approximately $65\%$ \citep{Beskos2013a}. Alternatively, the parameters could be adaptively tuned \citep{Wang2013a}; for this paper, we use the popular NUTS sampler \cite{Hoffman2014}, which tunes the trajectory length $L$ to avoid the sampler doubling back on itself. The HMC algorithm can be efficiently implemented using the popular STAN \citep{stan} software package. The STAN modelling language automatically tunes the HMC algorithm, and by using efficient automatic differentiation, the user need only express their posterior model.

\subsection{Importance sampling}
\label{sect.importance.basics}
A popular alternative to MCMC for estimating posterior expectations is the importance sampler \citep{robert_and_casella:1999:mcsm}. Given a proposal density, $q(\theta)$, and an unnormalised posterior density,
$\pi(\theta)$, importance sampling \citep[e.g.][]{Geweke1989} aims to
estimate  expectations of some measurable function of interest, $h(\theta)$ by
sampling from $q$. The starting point is 
\begin{equation}
  \label{eq:2}
  \mathbb{E}_{\pi/Z}[h(\theta)] = \frac{1}{Z}\int h(\theta)
  \frac{\pi(\theta)}{q(\theta)} q(\theta) d\theta  =
 \frac{1}{Z}\mathbb{E}_q[h(\theta)\overline{w}(\theta)], 
\end{equation}
where $\overline{w}(\theta):=\pi(\theta)/q(\theta)$ and $Z:=\int
\pi(\theta) \md\theta$ is the
normalisation constant.
 
Consider a sequence, $\{\theta_i\}_{i=1}^\infty$ with marginal
density $q$. Provided that a strong law of large numbers (SLLN) applies, setting
$h(\theta)=1$ in the above equation implies that
$\hat{Z}_N:=\frac{1}{N}\sum_{i=1}^N\overline{w}(\theta_i)\rightarrow
Z$, almost surely, and hence, almost surely,
\begin{equation}
\label{eqn.importance.cvg}
\hat{E}_N(h):=
\frac{1}{N}\sum_{i=1}^Nw_N(\theta_i)h(\theta_i)\rightarrow \mathbb{E}_{\pi/Z}[h(\theta)].
\end{equation}
where $w_N(\theta):=\overline{w}(\theta)/\hat{Z}_N$. In Section \ref{sec:three-methods} we will use importance sampling to estimate expectations with respect to the combined posterior distribution.

\section{A Gaussian-process approximation to the posterior}
\label{sec:gaussian-processes}

\subsection{Gaussian-process approximations to the subposteriors}

Parallelising the MCMC procedure over $C$ computing nodes results in
$C$ subposteriors $\{\pi_c(\vartheta)\}_{c=1}^C$. The MCMC algorithm for each subposterior, $c$, has been iterated $J$ times to give $\mathcal{D}_{c} =
\{\vartheta_{j},\ell_c(\vartheta_j)\}_{j=1}^J$, where $\ell_c(\vartheta_j) = \log\pi_c(\vartheta_{j})$ and each pair consists of a sample from the Markov chain
with its associated log-subposterior density, up to some fixed additive
constant. We wish to convert this limited information on a finite set
of points to information about $\log \pi_c$ over the whole support of
$\vartheta$. We start, \textit{a priori}, by treating the whole log-subposterior (up to the same additive
constant), $\mathcal{L}_c(\vartheta)$, as random with a Gaussian-process 
prior distribution: 
\[
  \mathcal{L}_c(\vartheta) \sim \mathcal{GP}(m(\vartheta),K(\vartheta,\vartheta^\prime)),
\]
where $m: \vartheta \rightarrow \mathbb{R}$ and $K: \vartheta \times \vartheta
\rightarrow \mathbb{R}$ are, respectively, the mean and covariance
functions. We model $\log \pi_c(\vartheta)$ rather than $\pi_c(\vartheta)$, so that our
approximation to the overall log-posterior will be a sum of Gaussian-process (Section
\ref{sec:merg-subp}); modelling the log-posterior also avoids the need
for non-negativity constraints when fitting the GP\footnote{GaussianProcesses.jl \citep[][]{fairbrother2017}, is used for simulations.}.

The mean function and covariance function are chosen by
the user. A mean function of zero, $m(\vartheta)=0$, would be inappropriate in this
setting as our prior must be the logarithm of a probability density
function up to a \textit{finite} additive constant. We
ensure that $\int \exp\left\{\mathcal{L}_c(\vartheta)\right\} \md \vartheta <\infty$ almost surely through
 a quadratic mean function of the form 
\[
m(\vartheta) = \beta_0 + \vartheta_1^\top\beta_1 + \mbox{diag}(\vartheta^\top V^{-1}\vartheta)\beta_2,~~~\beta_2<0.
\]
Here $V$ is the empirical covariance the posterior for $\vartheta$ obtained from the
MCMC sample and $\beta_i,~(i=0,1,2)$ are unknown constants. See Section \ref{sec:illustration} for a discussion on the choice of mean function.

The covariance function $K(\cdot,\cdot)$, determines the smoothness of the log-subposterior, which we shall assume is continuous with respect to $\vartheta$. A popular choice is the squared-exponential function \citep[e.g.][]{rasmussen}
\begin{equation}
  \label{eq:SQkernel}
K(\vartheta,\vartheta^\prime) = \omega^2\exp\left(-\frac{1}{2}(\vartheta-\vartheta^\prime)^{T}\Lambda^{-1}(\vartheta-\vartheta^\prime)\right),  
\end{equation}
where $\Lambda$ is a diagonal matrix and $\omega$ are hyperparameters. In this
paper we analytically marginalise $\beta_0$ and $\beta_1$
\citep{OHagan1978} and estimate $\beta_2$ and the kernel
hyperparameters through maximum likelihood (details given in Chapter 5 of \citet{rasmussen}). Alternative functions can be applied and may be more appropriate depending on the characteristics of the log-subposterior density.

Given the choice of prior, $\mathcal{D}_{c}$ are observations of this Gaussian-process generated from an MCMC algorithm targeting the subposterior $\pi_c(\vartheta)$, giving up to a constant of proportionality the posterior distribution,
\begin{equation}
  p(\gp_c(\vartheta)|\mathcal{D}_{c}) \propto p(\mathcal{D}_{c}|\gp_c(\vartheta))p(\gp_c(\vartheta)).
\label{eq:9}
\end{equation}
Define
$\mathcal{L}_c(\vartheta_{1:J}) := 
\{\mathcal{L}_c(\vartheta_1),\ldots,\mathcal{L}_c(\vartheta_J)\}$ and,  
for some parameter, or parameter vector,
$\theta:=\theta_{1:N}:=(\theta_{1},\dots,\theta_{N})$, define 
$\mathcal{L}_c(\theta_{1:N}) :=
\{\mathcal{L}_c(\theta_{1}),\ldots,\mathcal{L}_c(\theta_{N})\}$. 
We require the
posterior distribution of $\mathcal{L}_c(\theta_{1:N})|\mathcal{D}_c$, that is the
conditional distribution
$\GP_c(\theta_{1:N})|\{\GP_c(\vartheta_{1:J})=\gp_c(\vartheta_{1:J})\}$. Since the joint distribution
between $\mathcal{L}_c(\vartheta_{1:J})$ and
$\mathcal{L}_c(\theta_{1:N})$ is multivariate Gaussian, the
conditional,
$\GP_c(\theta_{1:N})|\{\GP_c(\vartheta_{1:J})=\gp_c(\vartheta_{1:J}\})$ is
also multivariate Gaussian, 
\begin{equation}
 \mathcal{L}_c(\theta_{1:N})|\mathcal{D}_c \sim \mathcal{N}(\mu_c(\theta_{1:N}),\Sigma_c(\theta_{1:N}))
\label{eq:10}
\end{equation}
with,
\begin{eqnarray}
  \label{eq:5}
  \mu_c(\theta_{1:N}) &=& m_c(\theta_{1:N}) + K^\top_{*}
  \tilde{K}^{-1}(\GP_c(\vartheta_{1:J})-m_c(\vartheta_{1:J})) \nonumber \\ 
  \Sigma_c(\theta_{1:N}) &=& K_{*,*} - K^\top_{*} \tilde{K}^{-1} K_{*},
\end{eqnarray}
and where $\tilde{K}=K(\vartheta_{1:J},\vartheta_{1:J})$, 
$K_{*,*} = K(\theta_{1:N},\theta_{1:N})$  and $K_*=K(\vartheta_{1:J},\theta_{1:N})$.

The posterior distribution for the GP,
$\mathcal{L}_c(\theta_{1:N})|\mathcal{D}_c$, up to a constant of proportionality, is a stochastic approximation of the log-subposterior
surface.

\subsection{Illustration}
\label{sec:illustration}
The Gaussian-process subposterior provides an estimate of the
uncertainty in the log-subposterior at points, $\theta$, where the log-subposterior has not been
evaluated. This contrasts with current approaches
\citep[e.g.][]{Scott2013,Neiswanger2013,Wang2013} to approximate the
subposterior which give no measure of uncertainty. This is
  illustrated below and used in Sections \ref{sec:gauss-proc-import} and \ref{sec:logistic-regression} to
  gauge the uncertainty in our estimates of posterior expectations.

\begin{figure}[h]
  \centering
   \subfigure{\includegraphics[width=0.32\textwidth,height=0.15\textheight]{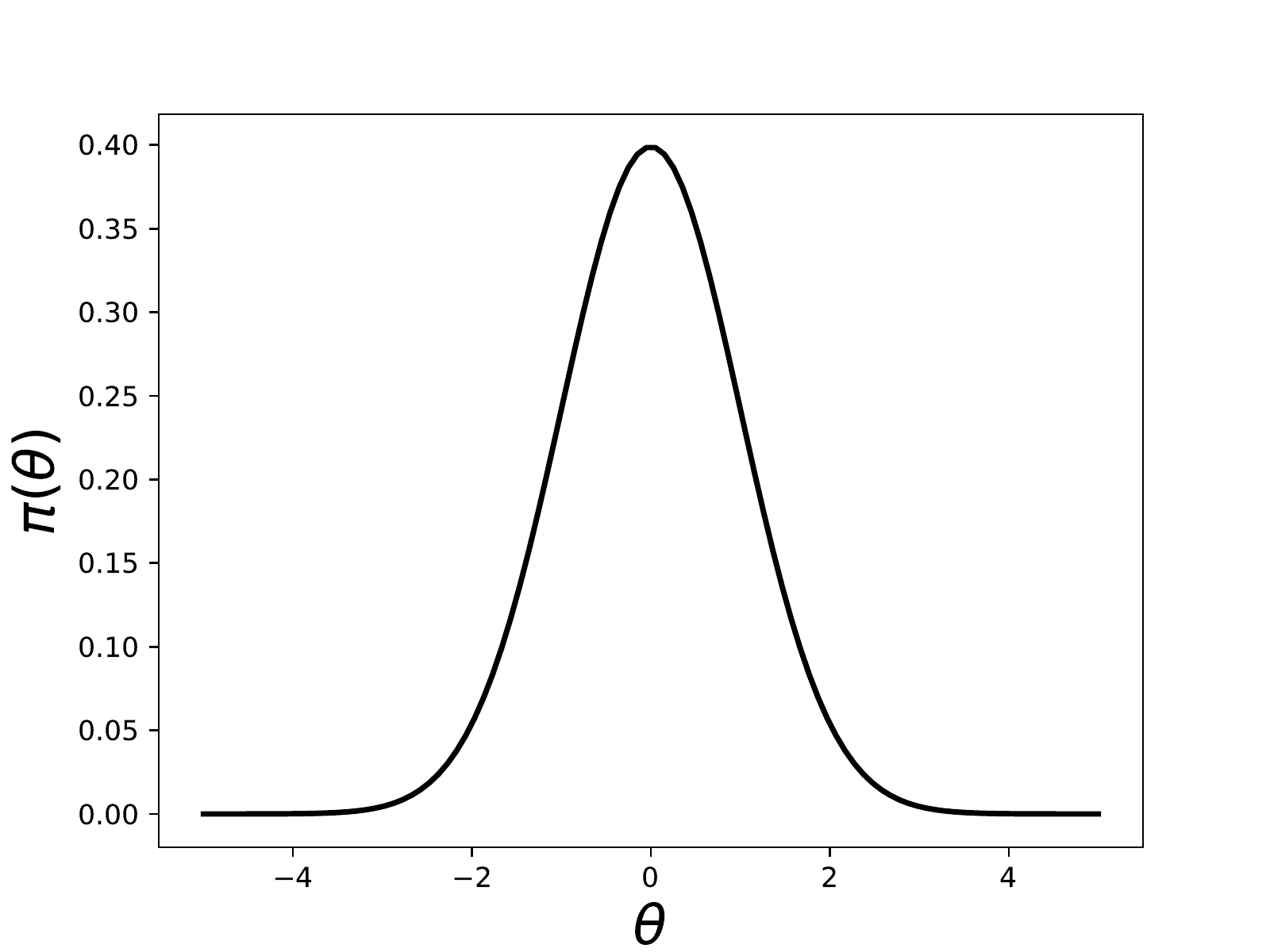}}
   \subfigure{\includegraphics[width=0.32\textwidth,height=0.15\textheight]{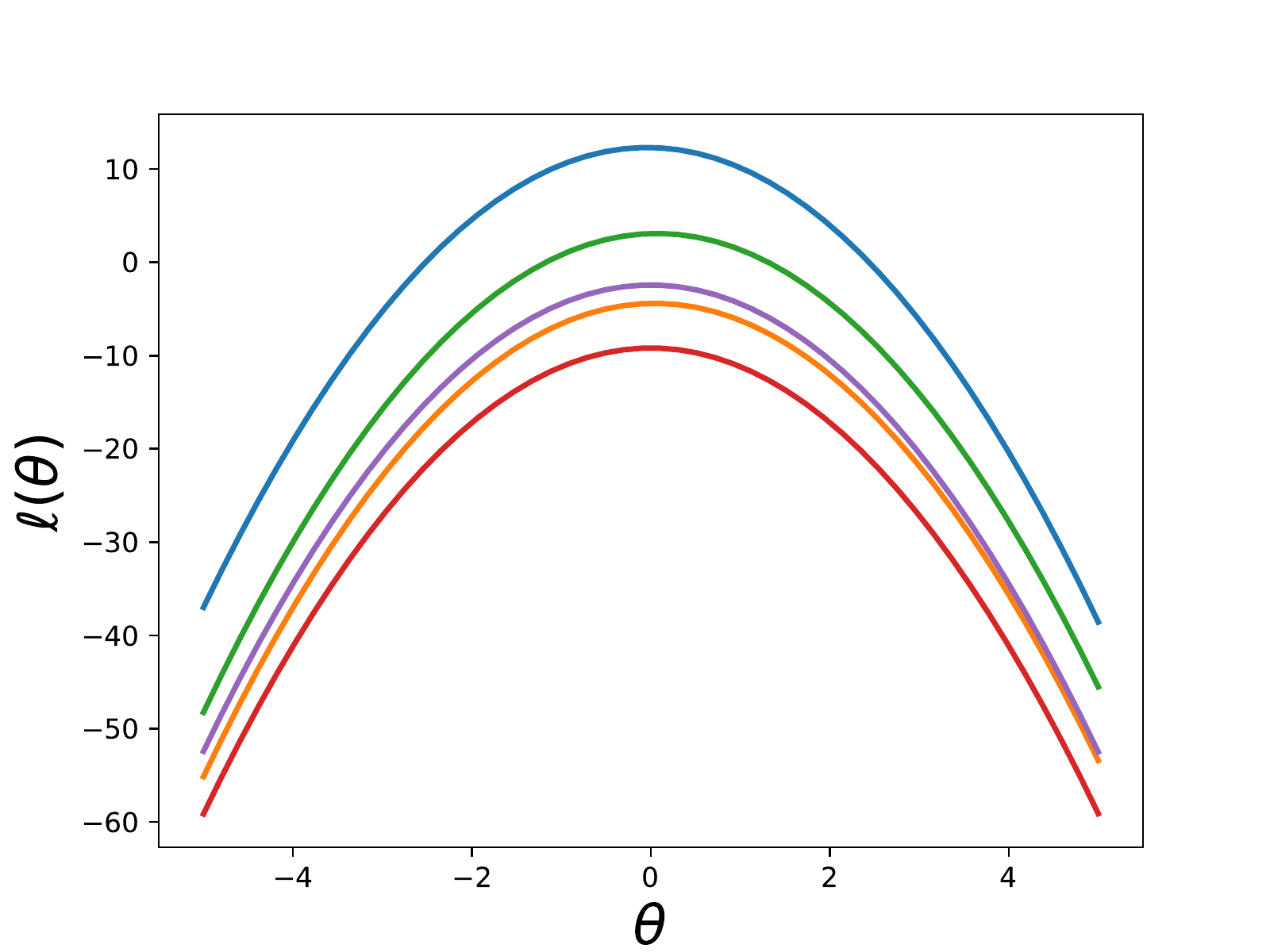}}
   \subfigure{\includegraphics[width=0.32\textwidth,height=0.15\textheight]{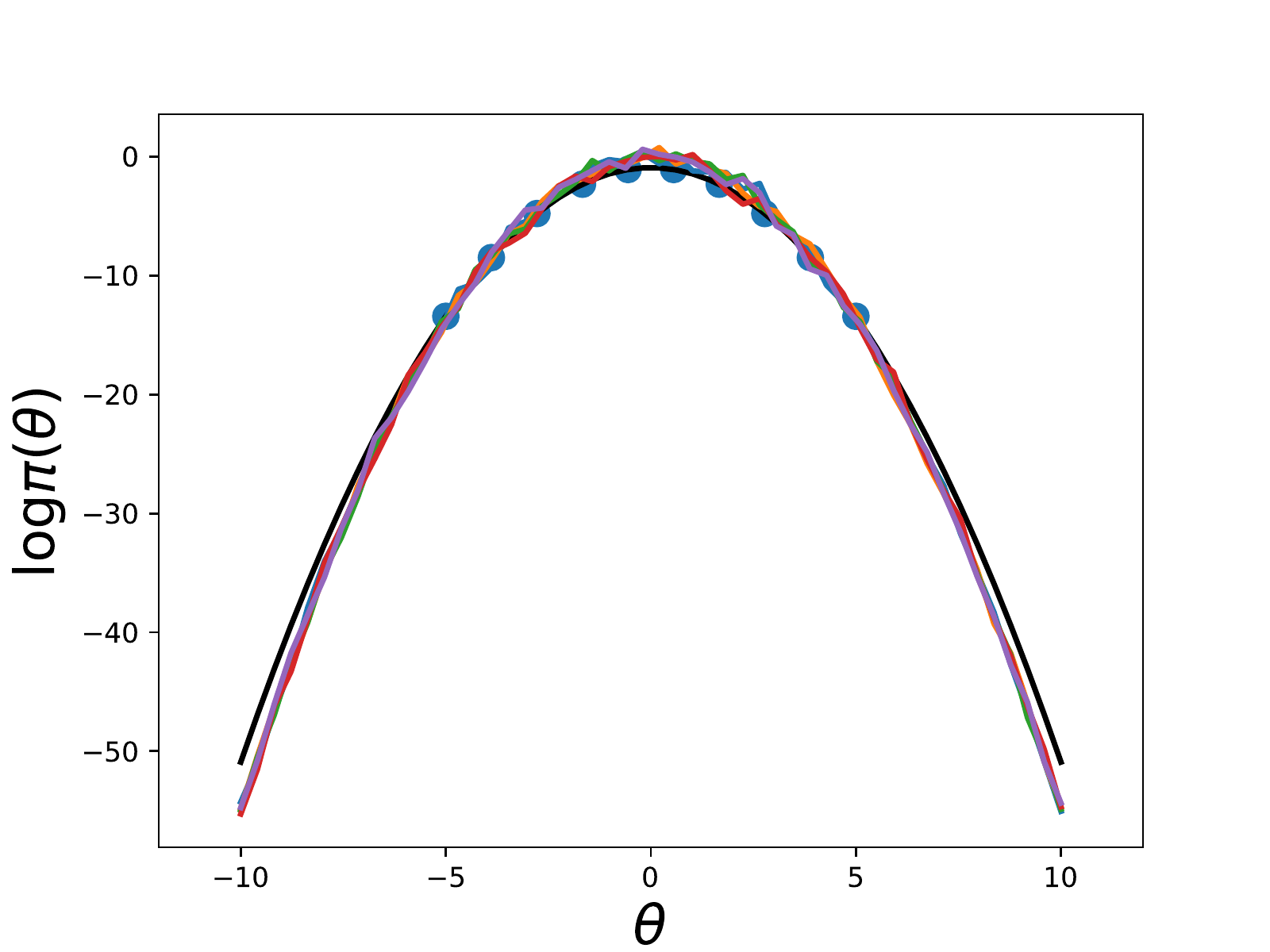}}
   \subfigure{\includegraphics[width=0.32\textwidth,height=0.15\textheight]{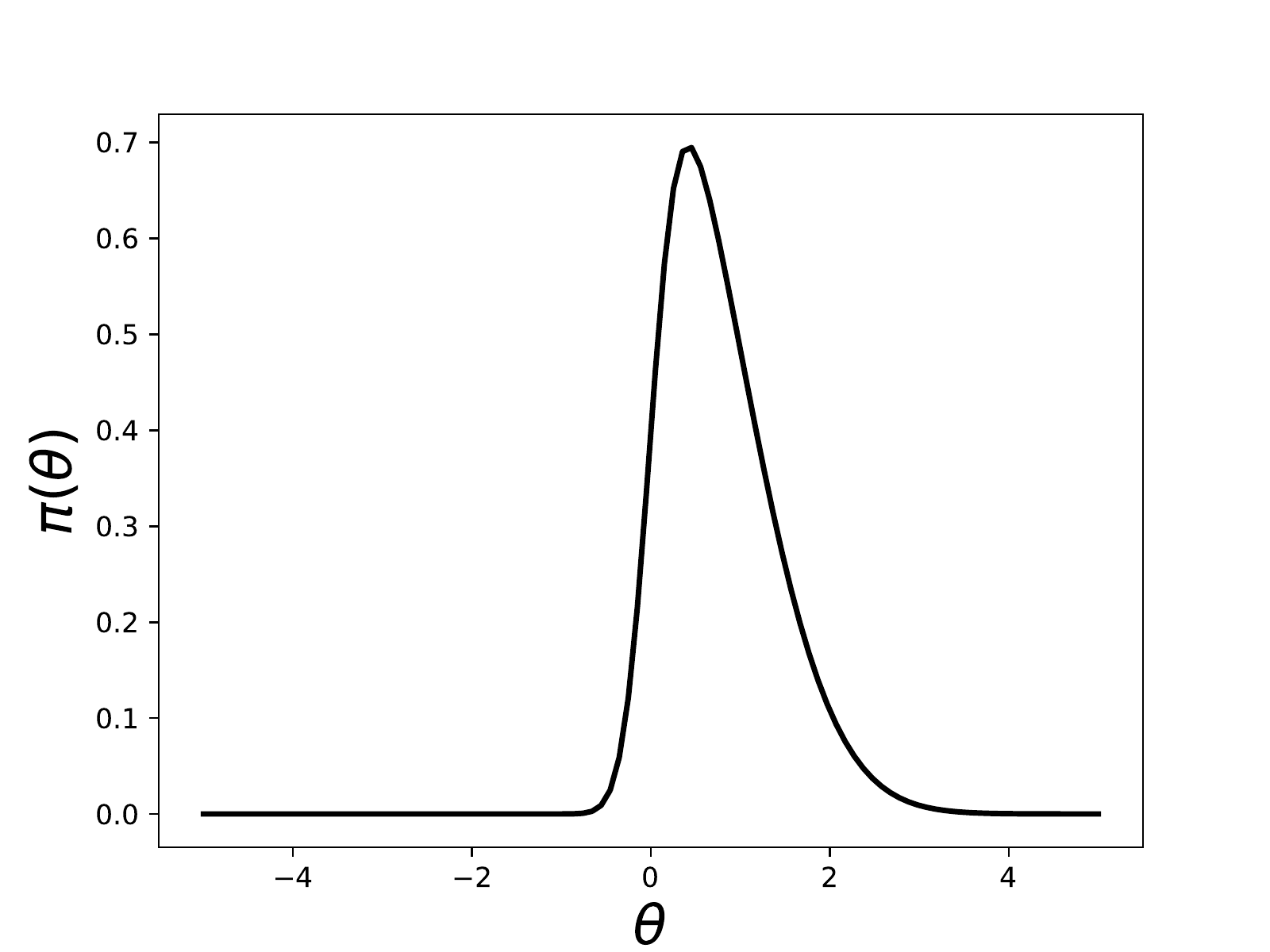}}
   \subfigure{\includegraphics[width=0.32\textwidth,height=0.15\textheight]{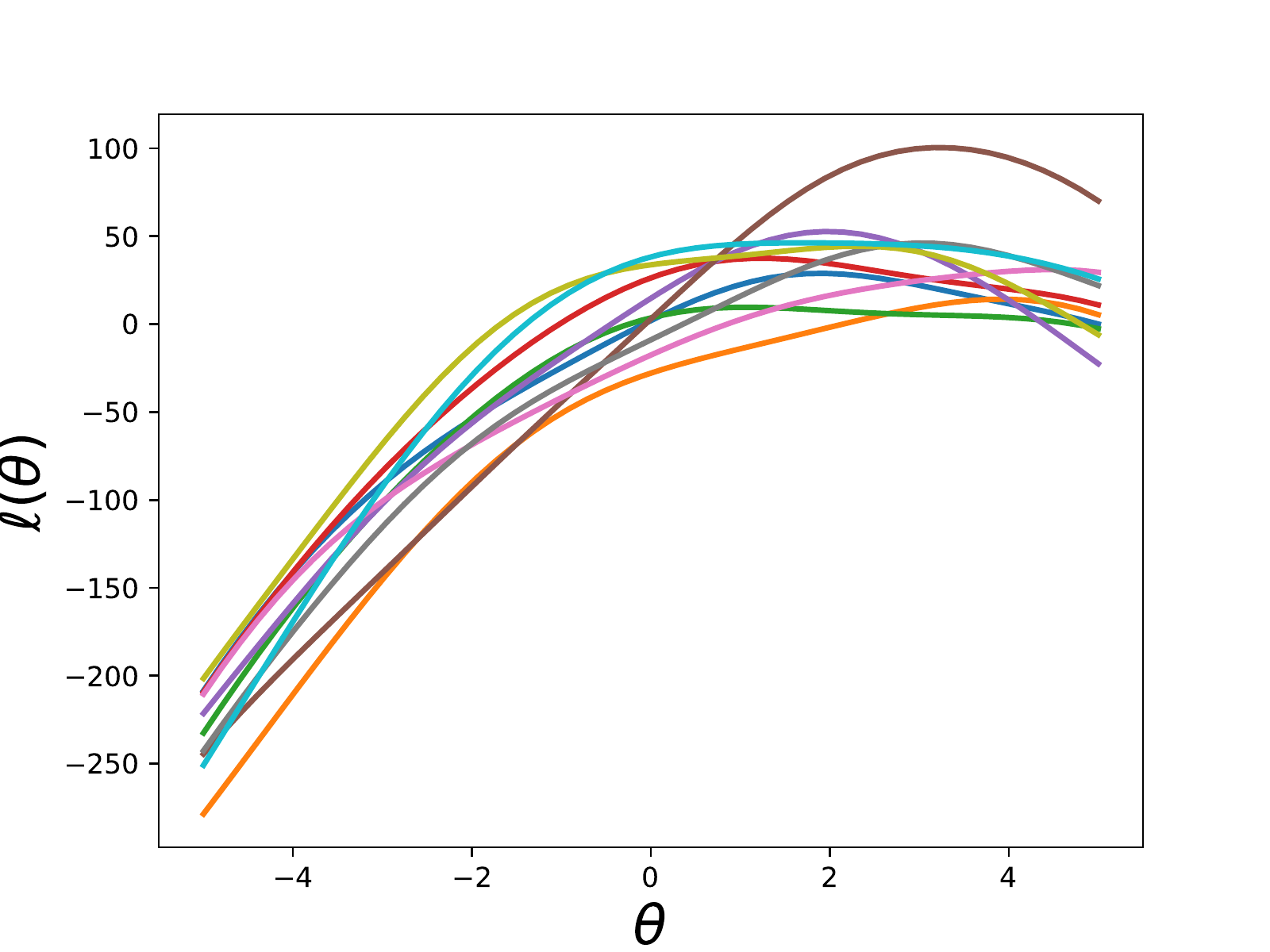}}
   \subfigure{\includegraphics[width=0.32\textwidth,height=0.15\textheight]{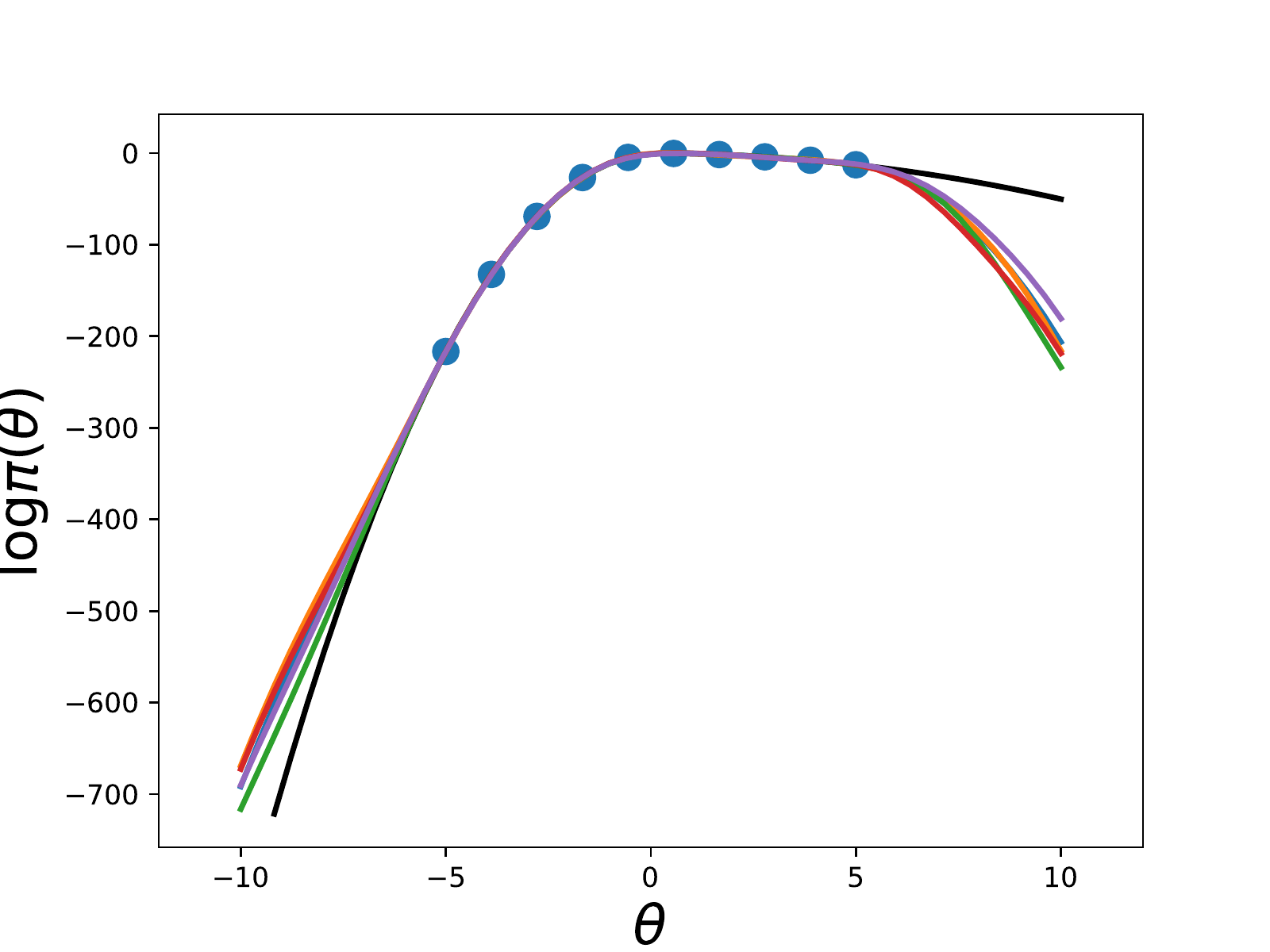}}
   \subfigure{\includegraphics[width=0.32\textwidth,height=0.15\textheight]{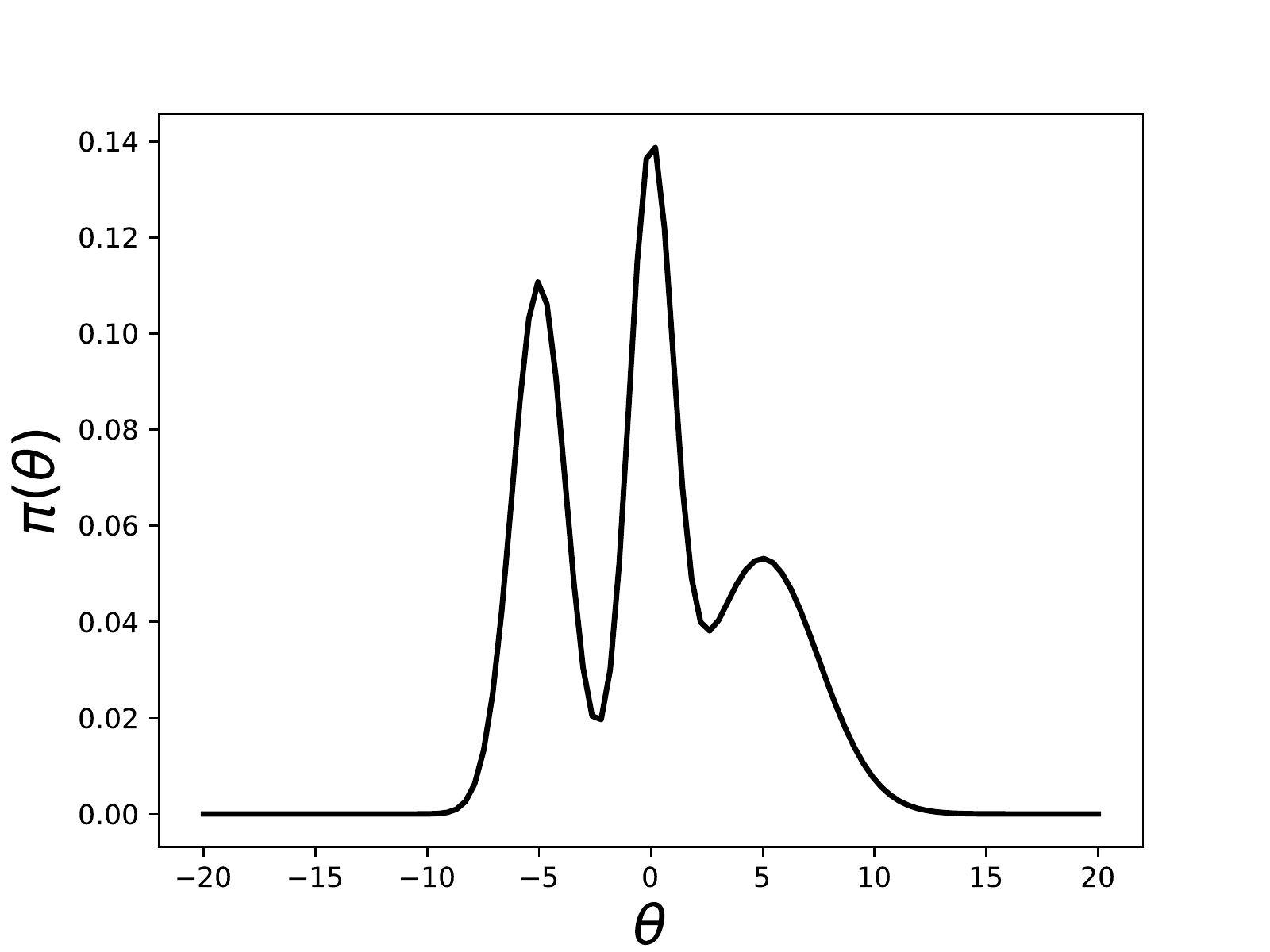}}
   \subfigure{\includegraphics[width=0.32\textwidth,height=0.15\textheight]{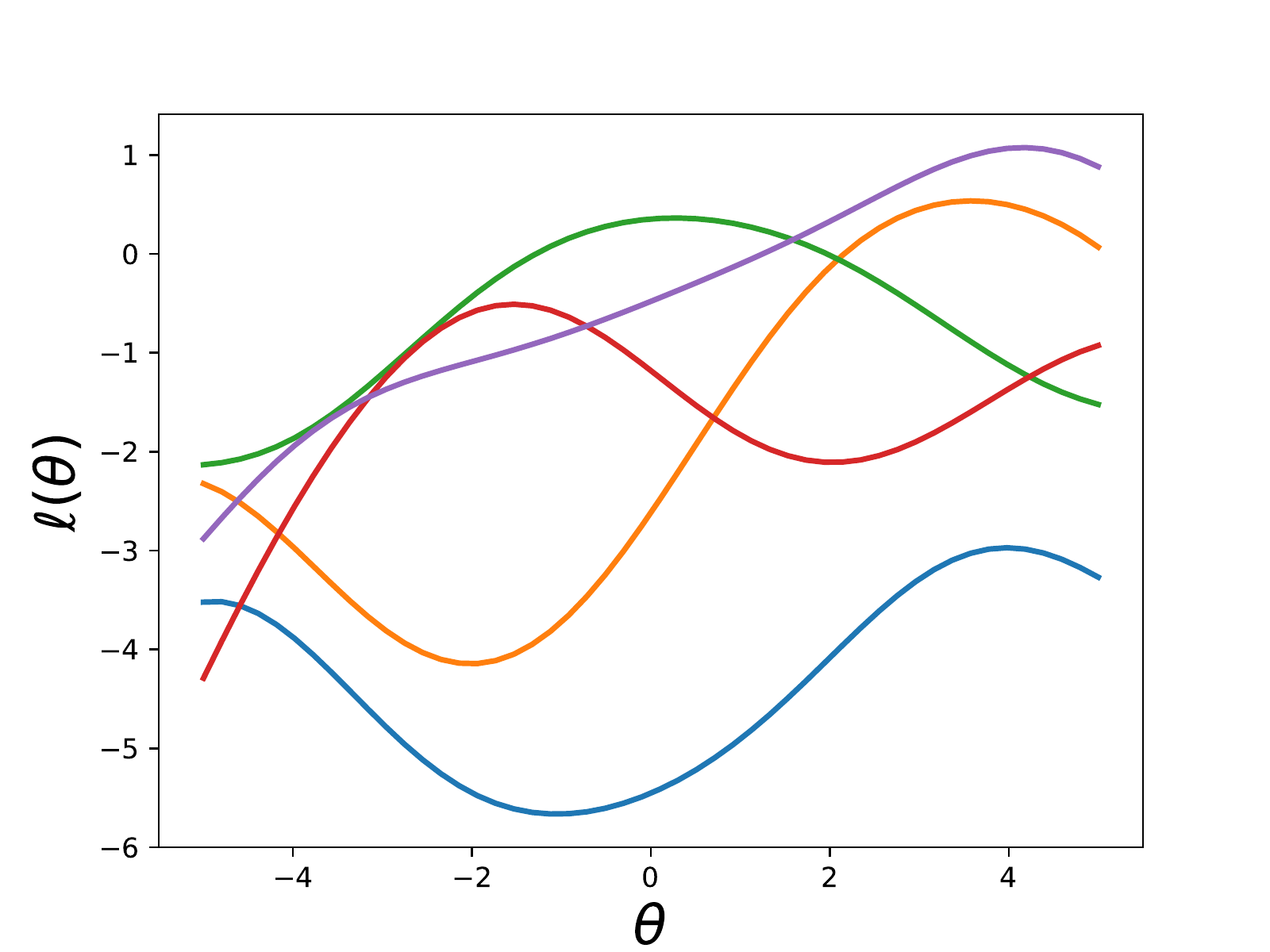}}
   \subfigure{\includegraphics[width=0.32\textwidth,height=0.15\textheight]{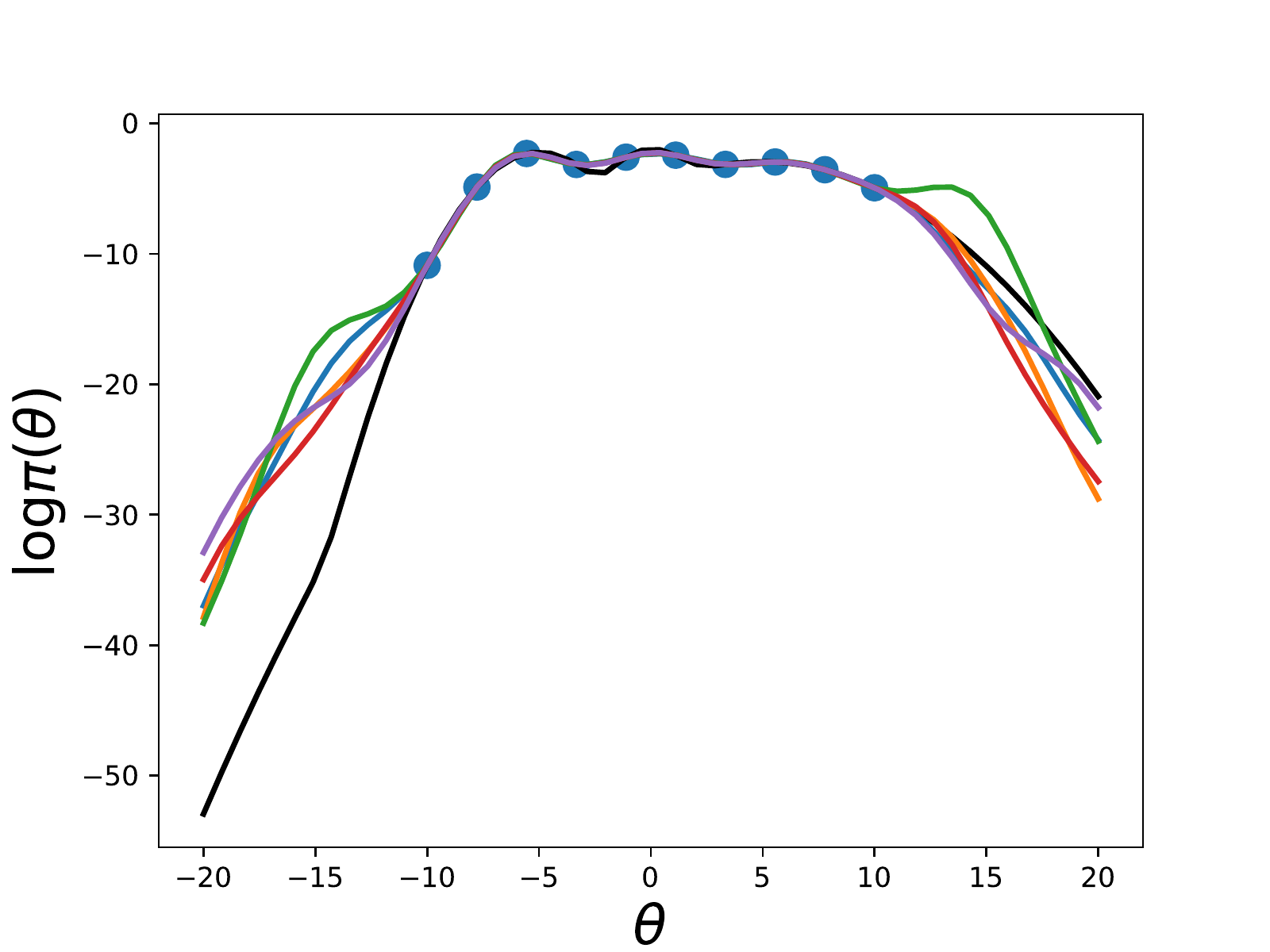}}
  \caption{Left: Target density. Centre: Five random samples from the GP prior with varying hyperparameters. Right: Posterior samples from the GP conditional on samples and log-density estimates of the target.}
  \label{fig:3gps}
\end{figure}

When approximating a log-density, we assume that the mean function $m_c(\theta)$ of the Gaussian process approximation to each log-subposterior, $\mathcal{L}_c(\theta)$, has a quadratic form. This assumption ensures that $\mathcal{L}_c(\theta) \rightarrow -\infty$ as $\theta \rightarrow \pm \infty$. Alternatively, the quadratic form of the log-subposterior could be modelled via the kernel by using a zero mean function and taking $K(\theta,\theta^\prime)$ to be the product of two linear kernels \citep{duvenaud2014automatic}. We test the quality of the proposed GP prior specification on three models shown in Figure \ref{fig:3gps}. Using our GP prior specification (\ref{eq:SQkernel}) we model the posterior of the Gaussian process approximation (\ref{eq:5}) on three models: standard Gaussian, skew Gaussian and mixture of Gaussians. Figure \ref{fig:3gps} gives the density of each model (left panel) along with a samples from the GP prior (centre panel), where the hyperparameters of the GP prior are estimated based on the observations, and the posterior (right panel) is fit to observations of the log-density of the respective model. The right panels of Figure \ref{fig:3gps} show that the GP provides a good approximation to the log-density of the respective models where the model has been evaluated. Outside of the range of evaluation points (i.e. as $\theta \rightarrow \pm \infty$) where there are no observations of the log-density, the GP reverts back to a quadratic form. We see that for these three models, the GP provides a good representation of the mass of the density, but can under-represent the tails of the distribution. Using our GP to model to approximate log-subposteriors can provide a good fit if the points at which we fit the GP are an appropriate representation of the true distribution. This will be the case if we fit the GP to MCMC samples $\{\vartheta_j\}_{j=1}^J$ drawn from the stationary distribution.

\subsection{Merging the subposteriors}
\label{sec:merg-subp}
Our next goal is to approximate the full posterior $\pi(\theta) \propto \prod_{c=1}^C \pi_c(\theta)$ by merging the subposteriors together. 
The approximation of each of the $C$ subposteriors as independent
Gaussian-processes, 
$\mathcal{L}_c(\theta) \sim \mathcal{GP}(\cdot,\cdot)~(c=1,\dots,C)$
leads directly to the approximation of the full log-posterior (up to
an additive constant) as the
sum of $C$ Gaussian-processes,
\begin{equation}
  \label{eq:fullGP}
\GP(\theta)|\mathcal{D} \propto \sum_{c=1}^C \left[\GP_c(\theta) |\mathcal{D}_c\right]=   
\mathcal{GP}\left(\sum_{c=1}^C \mu_c(\theta),\sum_{c=1}^C \Sigma_c(\theta)\right).
\end{equation}

Our assumption that the Gaussian-processes representing
the log-subposteriors $\{\GP_c\}_{c=1}^C$ are independent \textit{a
  priori} follows by assuming that the subposteriors are independent \eqref{eq:3}. This may not be true in practice where deviations from the
quadratic prior mean, $m_c(\vartheta)$, may be present across subposteriors. However, \textit{a posteriori} these deviations should be accounted for through the posterior mean
$\mu_c(\vartheta)$. Variability in the original partitioning of the data
into batches, and variability in the sample points, $\vartheta_{1:J}$,
across batches will both contribute to the more subtle variations of the GPs about their
individual posterior means, so that the posterior correlation should
be much smaller than the prior correlation.

\section{Approximating the full posterior}
\label{sec:three-methods}
We now detail three methods for approximating posterior expectations,
all of which utilise our Gaussian-process approximation to the full
posterior density. 

\subsection{The expected posterior density}
\label{sec:hamilt-monte-carlo}
Here we approximate the full posterior density (up to an unknown
normalising constant) by its expectation under the Gaussian-process approximation:
\begin{equation}
\label{eqn.approxE}
\pihat_E(\theta)\propto\mathbb{E}[\exp{(\GP(\theta)|\mathcal{D})}]
=\exp\left\{
\sum_{c=1}^C \mu_c(\theta) + \frac{1}{2}\Sigma_c(\theta)
\right\},
\end{equation}
using the properties of the log-Normal distribution. If the individual
GPs provide a good approximation to the individual log-subposteriors,
then $\Expect{\GP(\theta)}$ will be a good approximation to
the full log-posterior. 

The HMC algorithm then provides an efficient mechanism
for obtaining an approximate sample, $\{\theta_i\}_{i=1}^N$ from
$\pihat_E$. 
Evaluating the GP approximation at each iteration of this MCMC
algorithm is significantly faster than evaluating the true full
posterior, $\pi(\theta)$, directly. As is apparent from the leapfrog dynamics, HMC requires the gradient of $\log \pi_E$, and here the
tractability of our approximation is invaluable, since
\begin{eqnarray*}
\nabla\log \pihat_E &=& \sum_{c=1}^C \frac{\partial}{\partial \theta}
\mu_c(\theta) + \frac{1}{2}\frac{\partial }{\partial \theta}
\Sigma_c(\theta) \\
&=& 
\sum_{c=1}^C \frac{\partial}{\partial \theta} m(\theta) 
+\frac{\partial K_*^\top}{\partial
  \theta}\tilde{K}^{-1}(\ell_c(\vartheta_{1:J})-m(\vartheta_{1:J})) 
+\frac{1}{2}  \frac{\partial }{\partial \theta}K_{*,*} - \frac{\partial }{\partial \theta}K^\top_{*} \tilde{K}^{-1}.
\end{eqnarray*}
Given a sufficiently large sample from $\pihat_E$, approximations of
posterior expectations can be highly accurate if
the individual GPs provide a good approximation to the
log-subposteriors. Moreover, the approximation $\pihat_E(\theta)$ to the full posterior can be further improved by using importance sampling on the true posterior.

\subsection{Distributed importance sampling}
\label{sec:distr-import-sampl}

Unlike the proposal, $q$, in Section \ref{sect.importance.basics},
samples generated from the HMC algorithm represent an approximate, correlated
sample from an approximation to the true posterior, instead of exact,
independent samples from an approximation. Nonetheless, we may still correct for inaccuracies in
$\pihat_E$ using importance sampling while spreading the computational burden
across all $C$ cores.

The full sample from the HMC algorithm targeting $\pihat_E$,
$\{\theta_i\}_{i=1}^N$,
 is sent to each of the $C$ cores. Each of the $C$ cores then evaluates the true subposterior at each $\theta_i$. A single core
then combines the subposterior densities for each $\theta_i$  to provide
the full true posterior density: $\pi(\theta_i)=\prod_{c=1}^C\pi_c(\theta_i)$. To be clear, each
sub-posterior is evaluated at the \textit{same} set of $\theta$ values, allowing
them to be combined exactly. In contrast, the original HMC runs, performed on each individual
subposterior, created a different set of $\theta$ values for each
subposterior so that a straightforward combination was not possible. 

Each value from the sample, $\theta_i$, is then associated with an
unnormalised weight, $\overline{w}(\theta_i)=\pi(\theta_i)/\pihat_E(\theta_i)$. Defining
$\hat{Z}_N$ and $w_N(\theta)$ as in Section
\ref{sect.importance.basics}
 provides an
approximation $\hat{E}_N(h)$ to $\Expects{\pi}{h(\theta)}$ as defined
in \eqref{eqn.importance.cvg}.

Since the unknown normalising constants for both $\pi$ and $\pihat_E$
appear in both the numerator and the denominator of this expression,
they are not needed. Almost sure convergence of $\hat{E}_N(h)$ to $\Expects{\pi}{h(\theta)}$
as the HMC sample size, $N\rightarrow \infty$ relies on the strong law
of large numbers (SLLN) for
Markov chains \citep[e.g.][Theorem 4.3]{Tierney:1996}.
 In addition, if desired, an unweighted approximate sample
from $\pi$ may be obtained by resampling $\theta_i$ with
a probability proportional $w_i$. 

We expect our HMC importance proposal to be especially efficient
since it mimics the true posterior. However, other proposal
distributions based on competing algorithms for merging subposteriors
\citep[e.g.][]{Scott2013,Neiswanger2013,Wang2013} can be used instead;
these are compared in Section \ref{sec:simulation-study}. Algorithm
 \ref{alg:DIS} describes this general distributed importance sampler.

\begin{algorithm}[tb]
   \caption{Distributed Importance Sampler}
   \label{alg:DIS}
\begin{algorithmic}
  \STATE {\bfseries Input:} Proposal distribution $q(\theta_{1:N})$,
  which has marginal $q_1(\theta_i)$ identical for all $\theta_i,~i=1,\dots,n$. 
  \STATE - Sample $\theta_{1:N} \sim q(\cdot)$.
  \STATE - For $c=1,\ldots,C$ evaluate each subposterior $\pi_c(\theta_{1:N})$ .
  \STATE - Set $\pi(\theta_{i})=\prod_{c=1}^C \pi_c(\theta_{i})$.
\STATE - Weight the samples, $w_i = \frac{\pi(\theta_i)}{q_1(\theta_{i})}$.
  \STATE {\bfseries Output:} Weighted sample
  $\{w_i,\theta_i\}_{i=1}^N$ approximating $\pi(\theta)$.
\end{algorithmic}
\end{algorithm}

\subsection{Gaussian-process importance sampler (GP-IS)}
\label{sec:gauss-proc-import}
Finally, we present an importance sampler that uses
 the full posterior distribution of $\GP$, the GP approximation to the full
 unnormalised log-posterior conditional on $\{\vartheta_{c,j},\pi_c(\vartheta_{c,j})\}_{c=1,j=1}^{C,J}$. Compared with the importance sampler in Section
 \ref{sec:distr-import-sampl}, the set of points $\{\theta_i\}_{i=1}^N$ is generated
 from a simple proposal distribution, rather than the HMC algorithm applied to $\pihat_E$. Moreover, 
 given the set of points $\{\theta_i\}_{i=1}^N$ the
 computationally-expensive evaluation of each subposterior at this set
 of values is replaced with repeated, but relatively cheap sampling of 
 realisations of $\GP$ at these points. For a fixed number of GP training
 points, $J$, estimates of posterior expectations are no-longer
 asymptotically exact in $N$, however estimates of the uncertainty in
 these estimates are also supplied.

As in Sections \ref{sec:distr-import-sampl} and \ref{sect.importance.basics} we are interested in 
$\mathsf{I}_h:=\Expects{\pi}{h(\theta)}=\frac{1}{Z}\int\pi(\theta)h(\theta)\md
\theta$. 
Here we consider  approximating this with
\[
\mathsf{I}_h(\gp):=\frac{1}{Z(\gp)}
\int\exp\left\{\gp(\theta)\right\}h(\theta) \md \theta,
\]
where $\gp$ is a realisation of $\GP$ from the distribution in
\eqref{eq:fullGP} and $Z(\gp):=\int
\exp\{\gp(\theta)\}\md \theta$ is the associated normalisation
constant.

First, consider the hypothetical scenario where it is possible to store
$\ell$ completely and evaluate $\mathsf{I}_h(\ell)$. 
A set of $M$ realisations of $\GP$,
 $\{\gp_m\}_{m=1}^M$ would lead to $M$ associated estimates of
$\mathsf{I}_h$, $\{\mathsf{I}_h(\gp_m)\}_{m=1}^M$, which would approximate
the posterior distribution of $\mathsf{I}_h$ under \eqref{eq:fullGP}. 
The mean of these would then target,
the posterior expectation,
\[
\mathsf{I}_h^{\mathbb{E}}:=\mathbb{E}\left[
\frac{1}{Z(\GP)}\int h(\theta)\exp\left(\GP(\theta)\right)\md \theta\right].
\]
As an alternative, robust, point estimate, the median of
$\{\mathsf{I}_h(\gp_m)\}_{m=1}^M$ would target the posterior
median. Other posterior summaries for $\mathsf{I}_h$, such as a $95\%$ credible interval,
could also be estimated from the sample.

Unfortunately, it is not possible to store the infinite-dimensional
object, $\gp$; and even if it were, for moderate dimensions,
numerical evaluation of $\mathsf{I}_h(\ell)$ would be computationally
infeasible. Instead, we use importance sampling. 
Consider a proposal distribution
$q(\theta)$ that approximately mimics the true posterior distribution,
$\pi(\theta)$ and sample $N$ independent points from it: $\theta_{1:N}:=(\theta_1,\dots,\theta_N)$.
For each $m \in \{1,\dots,M\}$ we then sample the
\textit{finite-dimensional} object 
$(\gp_m(\theta_1),\dots,\gp_m(\theta_N))$ from the joint
distribution of the GP in \eqref{eq:fullGP}. For each such realisation 
we then construct
an approximation to the normalisation constant and to $\mathsf{I}_h(\gp)$:
\[
\hat{Z}(\gp_m):=
\frac{1}{N}\sum_{i=1}^N\overline{w}(\theta_i;\gp_m)
~~~\mbox{and}~~~
\hat{\mathsf{I}}_h(\gp_m):=\frac{1}{N\hat{Z}(\gp_m)}\sum_{i=1}^N\overline{w}(\theta_i;\gp_m)h(\theta_i),
\]
where
$\overline{w}(\theta;\gp):=\exp\{\gp(\theta)\}/q(\theta)$. 
The set $\{\hat{\mathsf{I}}_h(\gp_m)\}_{m=1}^M$ is then used in place
of $\{\mathsf{I}_h(\gp_m)\}_{m=1}^M$ for
posterior inference on $\mathsf{I}_h$.

For the specific case of $\mathsf{I}^{\mathbb{E}}_h$ a simplified
expression for the approximation may be derived:
\begin{equation}
\hat{\mathsf{I}}^{\mathbb{E}}_h
=
\frac{1}{N}\sum_{i=1}^Nw_ih(\theta_i),
  \label{eq:weights}
~~\mbox{where}~w_i:=\frac{1}{Mq(\theta_i)}\sum_{m=1}^M\frac{\exp\{\gp_m(\theta_i)\}}{\hat{Z}(\gp_m)}.  
\end{equation}
Algorithm \ref{alg:IS} creates point estimates based upon this.

The proposal density 
$q(\theta_i)$ should be a good approximation to the posterior
density. To create a computationally cheap
proposal, and with a similar motivation to the consensus Monte Carlo
approximation \citep{Scott2013}, we make $q(\theta_i)$ a multivariate
Student-t distribution on $5$ degrees of freedom
with mean and variance matching those of the Gaussian posterior that would arise given the mean and variance
of each subposterior and if each sub-posterior
were Gaussian, Alternatively, it would be possible to use the output from the HMC algorithm
of Section \ref{sec:hamilt-monte-carlo} in an analogous manner to the
way it is used 
in Section \ref{sec:distr-import-sampl}.

\begin{algorithm}[tb]
   \caption{GP Importance Sampler}
   \label{alg:IS}
\begin{algorithmic}
  \STATE {\bfseries Input:} GP approximation $\mathcal{L}(\theta)$ and proposal distribution $\mathsf{q}(\theta)$. 
  \STATE - Sample $\theta_{1:N} \sim \mathsf{q}(\cdot)$ iid.
  \STATE - Sample $m=1,\ldots,M$ realisations of the GP approximation.
  to the log-posterior $\GP_m(\theta_{1:N})$ in \eqref{eq:fullGP}.
%  \sim \mathcal{GP}\left(\sum_{c=1}^C \mu_c(\theta_{1:N}),\sum_{c=1}^C \Sigma_c(\theta_{1:N})\right)$
  \STATE - Weight the samples according to \eqref{eq:weights}.
  \STATE {\bfseries Output:} Weighted sample
  $\{w_i,\theta_i\}_{i=1}^N$, approximately from the marginal of $\{\GP,\pi{(\theta|\GP)}\}$.
\end{algorithmic}
\end{algorithm}
Many aspects of our importance sampler can, if necessary, be
parallelised: in particular, calculating $\mu_c(\theta_{1:N})$ and
$\Sigma_c(\theta_{1:N})$, and then sampling $\gp_1,\dots,\gp_m$ and  
obtaining the sample $\{\hat{\mathsf{I}}_h(\gp_m)\}_{m=1}^m$.

\subsection{Computational cost}
\label{sec:computational-cost}

We briefly review some of the notation in the paper as a point of reference for this section.
\begin{itemize}
\item $n$ := Number of data points, $y$.
\item $C$ := Number of processing cores (i.e. number of batches).
\item $J$ := Number of MCMC samples drawn from $\pi_c(\theta),~c=1,\dots,C$; for
  simplicity, we assume $J$ samples are drawn from each subposterior.
\item $N$ := Number of samples drawn from the approximation to the merged
  posterior $\hat{\pi}_E(\theta)$, or, for GP-IS, from the Student-t proposal.
\end{itemize}
The overall computational cost of applying the  methods in Sections
\ref{sec:hamilt-monte-carlo} and \ref{sec:distr-import-sampl} to
create an approximate (weighted) sample from the full posterior can be
summarised in three (four) steps:

\textbf{- Run MCMC on each subposterior} (see Section
\ref{sec:bayes-infer-with}). This step is common to all divide-and-conquer MCMC
algorithms \citep[e.g.][]{Scott2013,Neiswanger2013,Wang2015} and has a
cost of $\mathcal{O}(Jn/C)$.

\textbf{- Fit GP to each subposterior} (see Section
\ref{sec:gaussian-processes}). Fitting a Gaussian-process to each
subposterior has a cost of $\mathcal{O}(J^3)$ due to the
inversion of the $J\times J$ matrix $\tilde{K}$. One of the drawbacks of
Gaussian-processes is the computational cost. Faster, approximate
Gaussian-processes, referred to as \textit{sparse GPs}
\citep[e.g][]{Csat2002,Seeger2003,Quinonero-candela2005,Snelson2006}
can be used to reduce the computational cost (see Section \ref{sec:logistic-regression}). \footnote{Generally speaking,
sparse GPs introduce $p$ inducing points as training point locations
$\vartheta_{P_{i}},~i=1,\dots,p$, to fit the GP. The computational complexity of
such an approach is reduced if $P<<J$ to give an overall cost of $O(P^2J)$.} 
In this paper we apply the simpler speed-up technique of first
thinning the subposterior Markov chain; for example, using only every
twentieth sample. The thinned Markov chain has the same stationary
distribution as the full chain, but the autocorrelation 
is reduced and, more importantly for us, the sample contains fewer
points. Secondly, we remove duplicate samples from the subposterior; because we have the log-density of the subposterior, these duplicate samples provide no additional information when fitting the GP, and can cause the kernel matrix $\tilde{K}$ to become singular.  
Fitting $C$ independent GPs to each of the subposteriors is
embarrassingly parallel as the MCMC output from each subposterior is
stored on a separate core. 

\textbf{- Perform HMC on $\pihat_E$} (see Section
\ref{sec:hamilt-monte-carlo}). Each iteration of the HMC algorithm
requires an evaluation of $\mu_c$ and $\Sigma_c$ from \eqref{eq:5} with
$N=1$, and multiple evaluations of the gradient terms given in Section
\ref{sec:hamilt-monte-carlo}. Since $\tilde{K}^{-1}$ has already been
calculated, the total cost over all $N$ iterations of the HMC
algorithm is $\mathcal{O}(NJ^2)$. The cost of this step is equivalent to competing algorithms including \citep{Neiswanger2013,Wang2013}, which also use an MCMC-type step to sample from the approximation to the posterior.

\textbf{- Re-weight GP samples for DIS} (see Section \ref{sec:distr-import-sampl}). Our
 distributed importance sampler weights the approximate samples from
 $\pihat_E$ according to the true posterior and requires the
 subposteriors to be re-evaluated at each point in the GP-HMC
 sample. More generally, a sample from any sensible proposal
 distribution could be used. This has a cost of $\mathcal{O}(Nn/C)$.

\textbf{- GP-IS} (see Section \ref{sec:gauss-proc-import}). 
Taking the proposal, $q$, to be the Student-t distribution described
at the end of Section \ref{sec:gauss-proc-import},
creating a sample of size $N$ has a cost of $\mathcal{O}(N)$.
 For the sample, $\theta_{1:N}$, creation of each
$\Sigma_c(\theta_{1:N})~(c=1,\dots,C)$ in parallel is
 $\mathcal{O}(N^2J+NJ^2)$. Cholesky decomposition of their sum,
 $\Sigma$, is  
 $\mathcal{O}(N^3)$; however a spectral decomposition truncated to the
 largest $T$ eigenvalues is
 $\mathcal{O}(N^2T)$. The $M$ multiplications
 $\Sigma^{1/2}z$ (where $z$ is a vector of independent standard
 Gaussians) that generate realisations of $\GP$ can be spread between
 processors, leading to a cost of $\mathcal{O}(MN^2/C)$ (or
 $\mathcal{O}(MNT/C)$ for a truncated spectral decomposition).

GP-IS may, therefore, be preferable to DIS when $MN<<n$ (or
$MT<<n$).

\section{Experiments}
\label{sec:simulation-study}

In this section we compare our Gaussian-process algorithms for merging the subposteriors against several competing algorithms:
\begin{itemize}
\item \textbf{Consensus Monte Carlo} \citep{Scott2013}, where samples are weighted and aggregated.
\item \textbf{Nonparametric density product} \cite{Neiswanger2013}, where each subposterior is approximated using kernel density estimation.
\item \textbf{Semiparametric density product}\footnote{Implemented using the \textit{parallelMCMCcombine} R package} \cite{Neiswanger2013}, similar to the nonparametric method, but where subposteriors are approximated semiparametrically as in \cite{Hjort1995}.
\item \textbf{Weierstrass rejection sampler}\footnote{Implemented using the authors R package https://github.com/wwrechard/weierstrass} \citep {Wang2013}, where the nonparametric density estimates are passed through a Weierstrass transform to give the merged posterior.
\end{itemize}

We consider five examples (one in the Appendix) which capture common distributional features and popular statistical models: a warped Gaussian target, a mixture of bivariate Gaussians, a Bernoulli model with rare events which leads to a skewed posterior and two logistic regression models for large data sets. Additionally, in Appendix \ref{sec:multi-modal-distr}, we consider a mixture of Laplace distributions which only becomes identifiable with a large amount of data. These examples highlight some of the challenges faced by merging non-Gaussian subposteriors and the computational efficiency of large-scale Bayesian inference.

Our Gaussian-process approximation method is implemented using $J=100$
samples from the thinned chain for each subposterior to fit the GPs
for the Bernoulli and multimodal examples; for the logistic regression
examples $J=500$. Both for our methods and for competitor methods, $N=5000$ samples from each merged
posterior are created. To ensure a fair comparison, the sample from each
$\pihat_E$ that is used both directly and in our DIS algorithm is the
\textit{unthinned} output from the HMC run. The Student-t proposals
for the Gaussian-process importance sampler are iid. 

Weighted samples from DIS and GP-IS are converted to unweighted
samples by resampling with replacement, where the probability of
choosing a given $\theta$ is proportional to its weight.

For each of the models studied in this section we denote
the true parameter values by $\theta^*$ (when known). We obtain an accurate
estimate of the true posterior from a long MCMC run, thinned to
  a size of $N$, with samples
denoted by $\theta^f$  and the true posterior mean
and variance $\mathsf{m}_f$ and $\mathsf{V}_f$, respectively. Samples from the
approximation are denoted by $\theta^a$, and their 
mean and variance are $\mathsf{m}_a$
and $\mathsf{V}_a$. We use the following metrics to compare the competing methods:
\begin{itemize}
\item Mahalanobis distance, $D_{Mah.} = \sqrt{(\mathsf{m_a}-\mathsf{m_f})^\top \mathsf{V}_{\mathsf{f}}^{-1} (\mathsf{m_a}-\mathsf{m_f})}$.
\item Kullback-Leibler divergence for the Bernoulli and mixture example is calculated using a nearest neighbour search\footnote{Implemented using the \textit{FNN} R package} and for the logistic regression example, approximate multivariate Gaussian Kullback-Leibler divergence (see \citet{Wang2013} for details) between the true posterior $\pi$ and merged posterior $\hat{\pi}$ is calculated as
\[
D_{KL}(\hat{\pi}(\theta)||\pi(\theta)) = \frac{1}{2}(\mbox{tr}(\mathsf{V_f}^{-1}\mathsf{V_a})+(\mathsf{m_f}-\mathsf{m_a})^\top \mathsf{V_f}^{-1}(\mathsf{m_f}-\mathsf{m_a}) - d - \log(|\mathsf{V_a}|/|\mathsf{V_f}|))).
\]
\item Posterior concentration ratio, $\rho=\sqrt{\sum_{i=1}^N||\theta^{\mathsf{a}}_i-\theta^*||_2^2/\sum_{i=1}^N||\theta^{\mathsf{f}}_i-\theta^*||_2^2}$ \citep{Wang2015}, which gives a measure for the posterior spread around the true value $\theta^*$ ($\rho$=1 being ideal).
\item Mean absolute skew deviation, 
$\eta=\frac{1}{d}\sum_{i=1}^d  |\gamma^{\mathsf{a}}_i - \gamma^{\mathsf{f}}_i|$, where $i$ is the component,
  $\gamma_i=\mathbb{E}[\{(\theta_i-\mathsf{m}_{i})/\mathsf{V}^{1/2}_{ii}\}^3]$ is the
  third standardised moment,  and the superscripts $\mathsf{f}$ and
  $\mathsf{a}$ denote
  empirical approximations obtained from the samples obtained using
  the true 
  posterior and the approximation, respectively.
\end{itemize}

\subsection{Warped Gaussian model}
\label{sec:warp-gauss-post}

\begin{figure}[h]
  \centering
  \includegraphics[width=\textwidth]{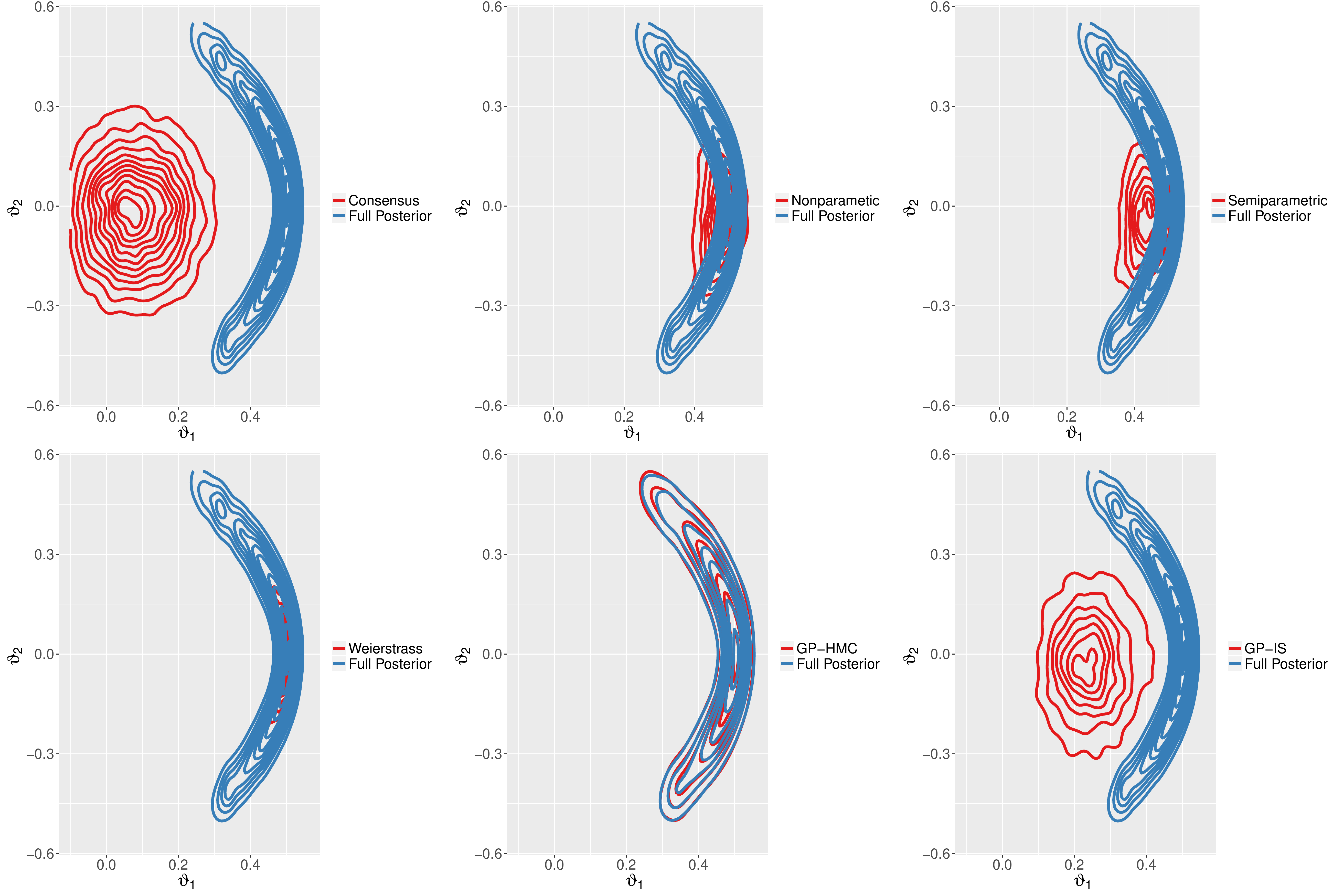}
  \label{fig:warped}
\caption{Posterior contour plots for the full posterior of the warped Gaussian target, overlayed with each of the competing algorithms.}
\end{figure}

We start by considering the warped Gaussian distribution, where the posterior, and subposteriors, exhibit a complex banana shape as a result of the lack of identifiability of the sign of one parameter. We simulate $n=50,000$ observations from a warped Gaussian distribution with density,
\[
p(y_i|\vartheta) = \mathcal{N}(y_i|\vartheta_1+\vartheta_2^2,\sigma^2),
\]
where $\vartheta = (0.5,0.0)$. We assume the variance $\sigma^2$ is known and the prior for $\vartheta$ is $\mathcal{N}(0,0.5)$. The data is split across $C=20$ processors with independent Hamiltonian MCMC algorithms \citep{stan} applied to each subset of the data targeting independent subposteriors. The subposteriors are re-merged using one of the competing methods and the posterior contour plots for each competing method are given in Figure \ref{fig:warped}. Out of the various re-merging algorithms, only the GP-HMC is able to accurately approximate the full posterior. The kernel density based methods are to reasonable capture the posterior mode and approximate posterior shape, but significantly underestimate the variance of $\vartheta_2$. The GP-IS sampler also struggles to adequately approximate the posterior. This is due to the importance proposal $q(\vartheta)$, which is a multivariate t-distribution approximation of the consensus Monte Carlo posterior. This example illustrates that the accuracy of the GP-IS sampler is highly dependent on the choice of importance proposal, and while the GP approximation can correct for some of the discrepancy in the proposal, generating a good approximation to the full posterior from the GP-IS sampler requires an importance proposal that sufficiently captures the posterior covariance structure.

\subsection{Mixture of Gaussians}
\label{sec:mixture-gaussians}

\begin{figure}[h]
  \centering
  \includegraphics[width=\textwidth]{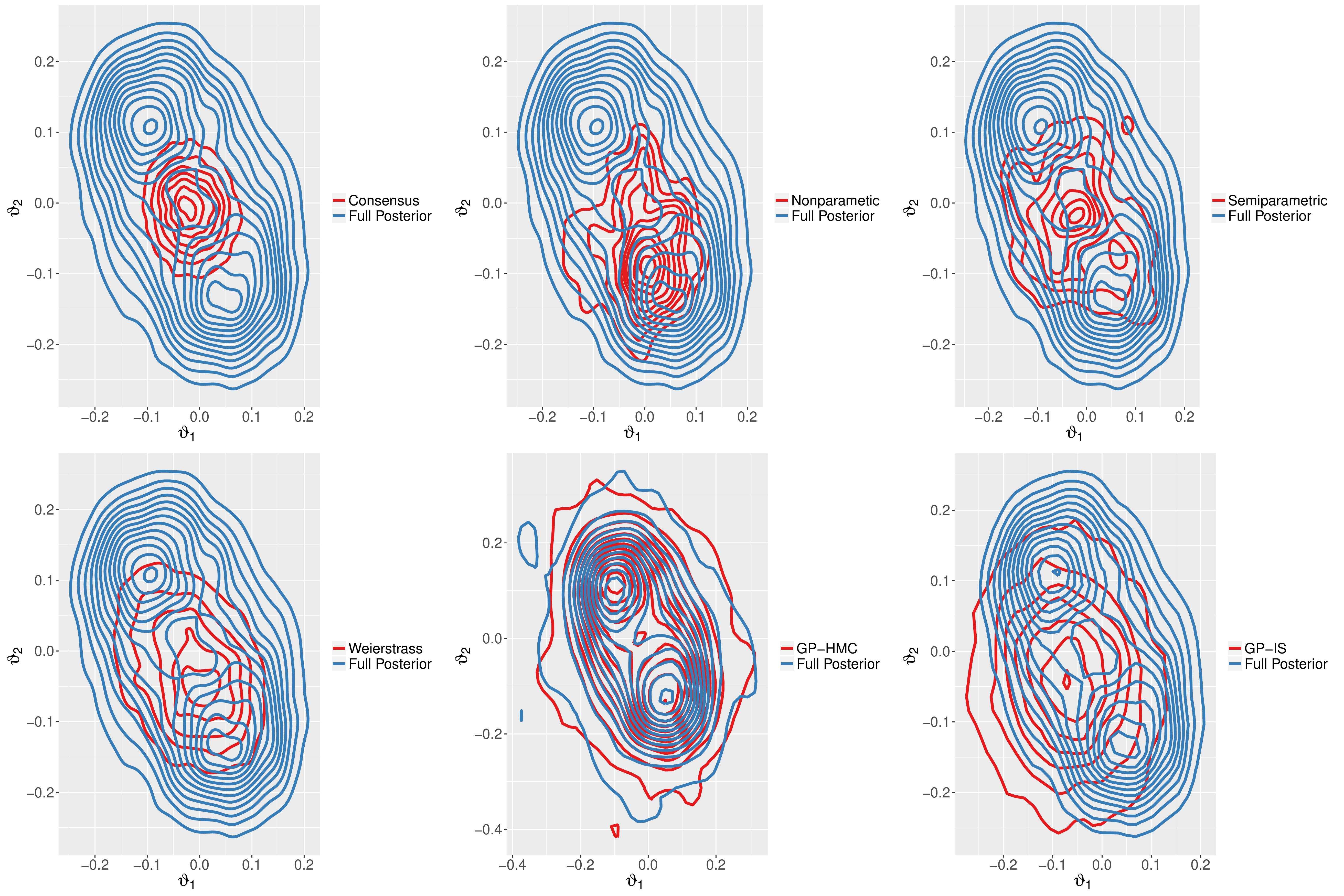}
  \label{fig:mixtures}
\caption{Posterior contour plots for the full posterior of the mixture of Gaussians target, overlayed with each of the competing algorithms.}
\end{figure}

Mixture models are popular in the divide-and-conquer MCMC literature \citep{Wang2013,Neiswanger2013}. We sample $n=50,000$ observations from a mixture of two bivariate Gaussian distributions with density,
\[
p(y_i|\vartheta) = \frac{1}{2}\mathcal{N}(y|\vartheta_1,\mathbf{I}_2) + \frac{1}{2}\mathcal{N}(y|\vartheta_2,\mathbf{I}_2),
\]
where $\vartheta_1 = (0.1,0.1)$ and $\vartheta_2=(-0.1,-0.1)$. We assume independent priors on $\vartheta \sim \mathcal{N}(0,100)$ and split the data across $C=20$ processors and run independent Hamiltonian MCMC \citep{stan} on each subposterior. This model has been constructed so that the posterior density exhibits bimodality. This is a result of placing the modes of the mixture components close together causing the MCMC algorithm to jump between the modes. Applying each of the merging algorithms to the subposteriors, we can see from the approximation to the full posterior, shown in Figure \ref{fig:mixtures}, that only the GP-HMC algorithm is able to sufficiently capture both the posterior mode and covariance structure. The kernel density methods are able to reasonably capture the mass of the posterior, but underestimate the covariance of the recombined full posterior. Unsurprisingly, the consensus Monte Carlo approximation is unable to capture the bimodality of the posterior, however, using the consensus approximation as a proposal within the GP-IS sampler, we are able to partially recover the shape of the posterior, but not sufficiently well to extract the posterior modes.

\subsection{Rare Bernoulli events}
\label{sec:rare-bern-events}

In the examples considered above, the subposteriors had approximately the same shape as the full posterior. This is not always the case and is largely dependent on how the data is split. It is possible that the data could be split in such a way some subposteriors are significantly more informative than others. We sample $n=10,000$ Bernoulli random variables, $y_i \sim \rm{Bern}(\vartheta)$, and assume a $\rm{Beta}(2,2)$ prior distribution for $\vartheta$. The data is split across $C=10$ processors. We set $\vartheta=C/n$ so that the probability of observing an event is rare. In fact, each subset only contains one success on average. Furthermore, we repeat this simulation study 100 times, each time randomly re-splitting the original data. By doing this we capture the uncertainty in our discrepancy metrics that result from the data splitting process.

\begin{figure}[t!]
  \centering
  \includegraphics[width=0.45\textwidth,height=0.25\textheight]{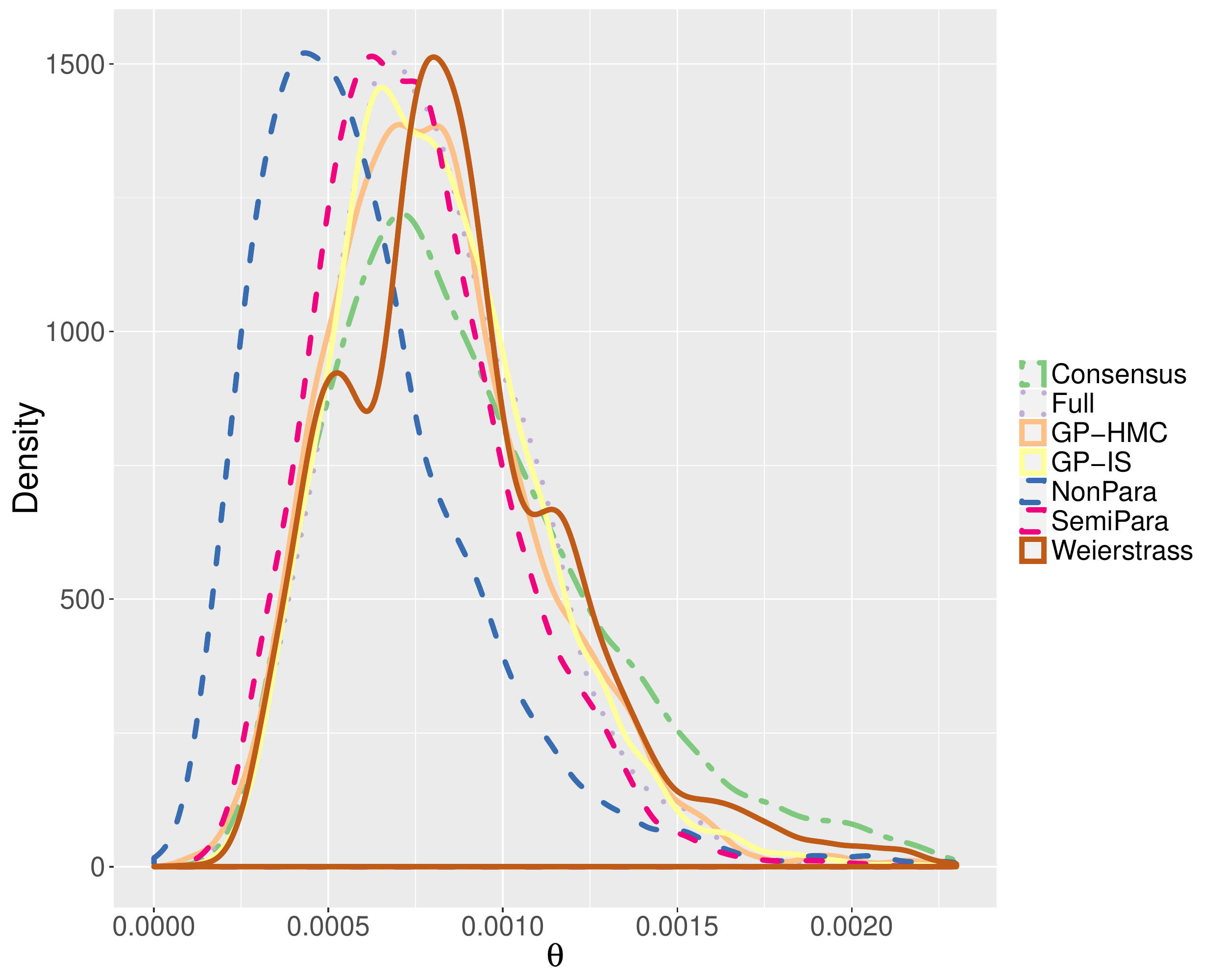}
  \includegraphics[width=0.45\textwidth,height=0.25\textheight]{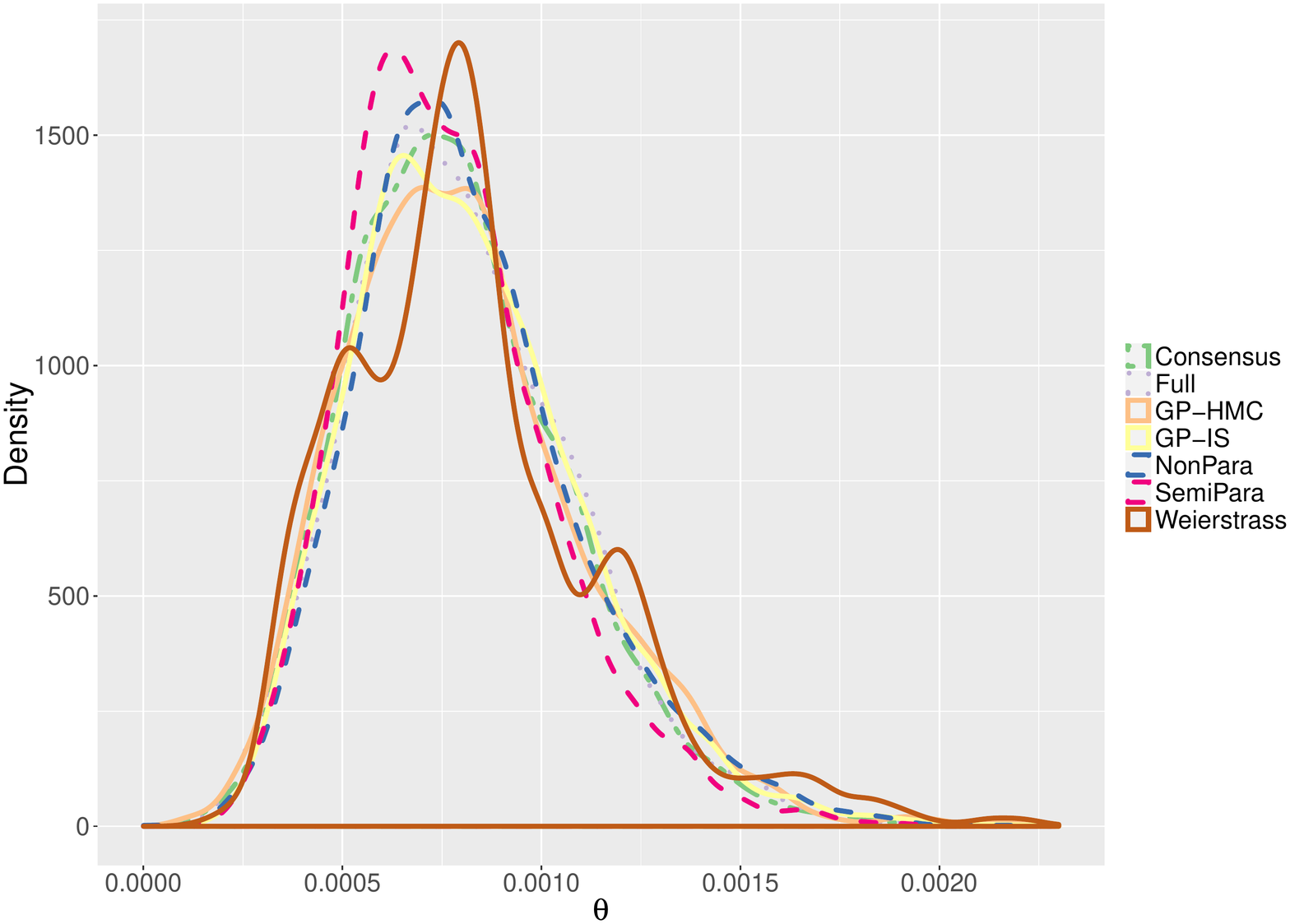}
  \caption{Left: Standard implementation of competing methods to approximate of the full posterior on a Bernoulli model (left). Right: Samples from each merging algorithm are used as a proposal in the distributed importance sampler}
  \label{fig:bernoulli}
\end{figure}

Figure \ref{fig:bernoulli} shows the posterior approximation resulting from each of the merging algorithms. Both GP-HMC and GP-IS samplers produce good approximations to the posterior. All of the competing algorithms can reasonably identify the mode of the posterior, but do not adequately fit the tail of the density. This example illustrates the advantage of the GP approximation, which utilises estimates of the log-subposterior density, over simply shifting and re-weighting subposterior samples using only the covariance of the subposteriors, as in the case of the consensus algorithm. 

We can generate samples from the full posterior using the distributed importance sampler (Alg. \ref{alg:DIS}), where samples from each of the merging algorithms can be used as a proposal. Figure \ref{fig:bernoulli} (right panel) shows that using the DIS improves the accuracy of all of the competing methods. This improvement is most noticeable for the consensus and nonparametric approximations. Ultimately, the overall accuracy of the approximation to the full posterior will dependent on the quality of the proposal distribution. 

Table \ref{tab:bernoulli} provides metrics to assess the accuracy of each of the merge algorithms. We report the mean and standard deviation (in brackets) of each metric taken over 100 simulations, where, for each simulation, we re-split the data. On average the GP-HMC and GP-IS samplers display the best performance across all metrics, most notably with regards to Kullback-Leibler divergence. The GP-HMC and GP-IS samplers also have the lowest standard deviation compared to the alternative subposterior merge algorithms. This improvement, however, comes at a higher computational cost than the competing methods. In Appendix \ref{sec:addit-results-rare} we further investigate the variability of the discrepancy metrics.

\begin{table}
\centering
\begin{tabular}{l|cccccc}
\hline
Algorithm & $D_{Mah.}$ &$D_{KL}(\pi||\hat{\pi})$ & $D_{KL}(\hat{\pi}||\pi)$ & $\rho$ & $\eta$ & Time\\
\hline
Consensus      &$1.69$ ($0.59$)&  $0.33$ ($0.33$) & $0.33$ ($0.40$) & $1.51$ ($0.55$) & $0.55$ $(0.18)$ & 0.07\\
Nonparametric  &$1.42$ ($0.16$)&  $0.40$ ($0.15$) & $0.46$ ($0.25$) & $1.20$ ($0.27$) & $1.00$ $(0.90)$ & 2.03\\
Semiparametric &$1.12$ ($0.25$)&  $0.27$ ($0.42$) & $0.28$ ($0.68$) & $\mathbf{1.03}$ ($0.28$) & $0.24$ $(0.43)$ & 2.80\\
Weierstrass    &$1.27$ ($0.25$)&  $0.20$ ($0.12$) & $0.14$ ($0.14$) & $1.26$ ($0.25$) & $0.14$ $(0.11)$ & 1.8\\
GP-HMC sampler &$\mathbf{1.03}$ ($0.06$)&  $\mathbf{0.09}$ ($0.02$) & $0.09$ ($0.02$) & $1.04$ ($0.07$) & $\mathbf{0.10}$ ($0.08$) & 13.08\\
GP-IS sampler  &$\mathbf{1.03}$ ($0.05$)&  $\mathbf{0.09}$ ($0.02$) & $\mathbf{0.08}$ ($0.02$) & $1.04$ ($0.06$) & $\mathbf{0.10}$ ($0.07$) & 14.22\\
\hline
\end{tabular}
\caption{Mean discrepancy of various merge algorithms over 100 simulated rare event Bernoulli models. Kullback-Leibler results are
  reported as ($\times 10^{1}$) and $D_{Mah.}$ as ($\times
  10^5$). Average execution time is given in seconds. Results in brackets represent standard deviation of metrics over 100 datasets. 
}
\label{tab:bernoulli}
\end{table}

\subsection{Logistic regression}
\label{sec:logistic-regression}

\textbf{Synthetic data set} We use a synthetic data set based on internet click rate behaviour, where one of the covariates is highly predictive, but rarely observed. The dataset has $n=10,000$, with $5$ covariates and is generated according to Section 4.3 of \citet{Scott2013}. The data are partitioned across $10$ machines. We repeat this experiment 100 times, randomly re-partitioning the original dataset for each experiment.

Additionally to the algorithms discussed at the start of Section \ref{sec:simulation-study}, we introduce sparse Gaussian process versions of the GP-HMC and GP-IS samplers. We apply the sparse GP presented by \cite{titsias2009variational}, which uses a variational approach to infer the inducing inputs (see \cite{Quinonero-candela2005} for a review of sparse GP approximations).

The posterior distribution for this model is approximately Gaussian and all algorithms perform equally well in this setting (see Table \ref{tab:logistic}). The standard deviation of the discrepancy metrics is generally lower than that of the Bernoulli model (Table \ref{tab:bernoulli}). For the logistic regression example, there is less variation in the distribution of the data batches over repeated simulations, compared to the Bernoulli example, where there is greater variability from splitting the data. The distributed importance sampler is applied to the posterior approximations with the results given in Appendix \ref{sec:logist-regr-model}. We also provide additional simulations where the posterior is approximated with varying sample sizes. We show that it is possible to apply our GP algorithms with fewer samples, giving a reduced computational cost, while maintaining a high level of accuracy.

\begin{table}
\centering
\begin{tabular}{l|cccccc}
\hline
Algorithm & $D_{Mah.}$ & $D_{KL}(\pi||\hat{\pi})$ & $D_{KL}(\hat{\pi}||\pi)$ &$\rho$ & $\eta$ & Time\\
\hline
Consensus      & 2.36 (0.07)          & \textbf{0.03} (0.01) & \textbf{0.03} (0.01) & 0.16 (0.00)          & 0.09 (0.03)          & 0.01 \\
Nonparametric  & 3.67 (0.25)          & 0.75 (0.14)          & 2.08 (0.54)          & 0.16 (0.00)          & 0.16 (0.06)          & 0.72 \\
Semiparametric & 2.28 (0.16)          & 0.23 (0.09)          & 0.17 (0.06)          & 0.16 (0.00)          & 0.17 (0.08)          & 4.23 \\
Weierstrass    & 2.37 (0.13)          & 0.11 (0.06)          & 0.10 (0.04)          & 0.16 (0.00)          & 0.17 (0.06)          & 0.17 \\
GP-HMC         & \textbf{2.10} (0.71) & 0.62 (0.23)          & 0.67 (0.12)          & \textbf{0.75} (0.02) & 0.16 (0.18)          & 261  \\
GP-HMC-Sparse  & 2.69 (0.79)          & 0.69 (0.26)          & 0.70 (0.13)          & 0.71 (0.02)          & 0.18 (0.17)          & 94   \\
GP-IS          & 3.42 (0.08)          & 1.86 (0.14)          & 2.85 (0.23)          & 0.74 (0.00)          & \textbf{0.05} (0.01) & 16.43\\
GP-IS-Sparse   & 3.42 (0.08)          & 1.85 (0.14)          & 2.86 (0.24)          & 0.72 (0.00)          & \textbf{0.05} (0.01) & 14.6 \\
\hline
\end{tabular}
\caption{Mean discrepancy of various merge algorithms over 100 data splits of the logistic regression model with simulated data. Average execution time is given in seconds. Results in brackets represent standard deviation of metrics over 100 data splits. }
\label{tab:logistic}
\end{table}

\textbf{Real data set} We conduct divide-and-conquer MCMC experiments on the Hepmass \footnote{https://archive.ics.uci.edu/ml/datasets/HEPMASS} data set. The challenge is to accurately classify the collisions of exotic particles by separating the particle-producing collisions from the background source. The full data set contains 10.5 millions instances with 28 attributes representing particle features. In our experiments, we use the first million instances and partition the data equally across $C=20$ machines. 

Table \ref{tab:logistic_hepmass} gives the mean and standard deviation of the discrepancy metrics for each algorithm taken over 100 simulations (additional plots given in Appendix \ref{sec:logist-regr-model-1}). For this example, the subposteriors and full posterior distributions are approximately Gaussian and so all methods approximate the full posterior with more or less the same level of accuracy. As discussed in \cite{Neiswanger2013}, nonparametric methods scale poorly with dimension (i.e. number of covariates) with the Weierstrass and semiparametric algorithms performing better than the simple nonparametric method. As a result, applying the DIS step does not lead to a significant improvement in the approximation.

\begin{table}
\centering
\begin{tabular}{l|cccccc}
\hline
Algorithm & $D_{Mah.}$ & $D_{KL}(\pi||\hat{\pi})$ & $D_{KL}(\hat{\pi}||\pi)$ & $\eta$ & Time\\
\hline
Consensus      & 5.21 (0.05)          & \textbf{11.55} (0.05) & 11.46 (0.03)          & 0.10 (0.01)          & 0.03 \\
Nonparametric  & 9.83 (0.23)          & 16.57 (0.32)          & 30.04 (2.00)          & 0.11 (0.02)          & 4.15 \\
Semiparametric & 5.22 (0.11)          & 13.67 (0.31)          & 12.66 (0.14)          & 0.11 (0.02)          & 13.36 \\
Weierstrass    & 5.39 (0.27)          & 15.83 (0.41)          & 13.48 (0.15)          & 0.13 (0.01)          & 0.74 \\
GP-HMC         & 5.35 (0.59)          & 21.78 (0.63)          & 11.08 (2.24)          & 0.10 (0.01)          & 283.56 \\
GP-HMC-Sparse  & \textbf{5.06} (0.36) & 21.79 (0.71)          & \textbf{10.34} (1.30) & 0.10 (0.01)          & 137.94 \\
GP-IS          & 5.96 (0.06)          & 16.76 (0.28)          & 15.61 (0.18)          & \textbf{0.09} (0.01) & 15.95 \\
GP-IS-Sparse   & 5.97 (0.08)          & 16.78 (0.29)          & 15.69 (0.29)          & \textbf{0.09} (0.01) & 14.25 \\
\hline
\end{tabular}
\caption{Mean discrepancy of various merge algorithms over 100 data splits of the logistic regression model with the Hepmass dataset. Average execution time is given in seconds. Results in brackets represent standard deviation of metrics over 100 data splits. }
\label{tab:logistic_hepmass}
\end{table}

The major difference in the results from Table \ref{tab:logistic_hepmass} is the computational time. We see that the GP-IS sampler has comparable computational cost to the nonparametric algorithms, but the GP-HMC samplers have the highest cost overall. It is important to note that, while more expensive than some cheaper competitors, the goal is to produce highly accurate posterior approximations that circumvent applying MCMC to the full dataset. For this example, running an HMC algorithm on the full data takes 19.4 hours. Therefore, relative to this computational cost, applying the GP-HMC sampler accounts for only $0.4\%$ of the total time.

Finally, in Section \ref{sec:gauss-proc-import}, we note that the GP-IS sampler draws multiple realisations from the posterior distribution of the GP approximation to the posterior. Each of these realisations provides an estimate of the expectation of interest, their centre (mean or median) provides a point estimate and their spread (2.5\% and 97.5\% quantiles) provide a measure of the uncertainty. In Table \ref{tab:gpis} we estimate the posterior mean and variance of two randomly selected parameters (for ease of presentation) and compare these estimates against those calculated from an MCMC run on the full posterior. Sampling $M=500$ realisations from the GP, we report the mean, median and $95\%$ interval for estimates of the mean and find that these results are consistent with those of the full applying MCMC on the full data. 
\begin{table}
\centering
\begin{tabular}{l|c|ccc}
\hline
 & $\pi(\vartheta)$ & Mean & Median & Quantiles $(2.5\%,97.5\%)$\\
\hline
$\Expects{\hat{\pi}}{\vartheta_1}$      & $0.45$& $0.45$ & $0.45$ & $(0.44,0.46)$\\
$\Vars{\hat{\pi}}{\vartheta_1}$ $(\times 10^5)$ & $1.26$ & $1.25$  & $1.24$ & $(1.21,1.28)$ \\
$\Expects{\hat{\pi}}{\vartheta_{17}}$ $(\times 10^2)$ & $0.22$ & $0.22$ & $0.21$ & $(0.17,0.28)$ \\
$\Vars{\hat{\pi}}{\vartheta_{17}}$  $(\times 10^5)$  & $7.79$ & $7.75$ &  $7.61$ & $(7.60,7.89)$ \\
\hline
\end{tabular}
\caption{Expectation and variance of $\vartheta_1$ and $\vartheta_{17}$ from the logistic regression model with the HEPMASS dataset. Mean, Median and quantile estimates of the quantities are calculated from $500$ samples from the GP-IS sampler (i.e. M=500).}
\label{tab:gpis}
\end{table}

\section{Discussion}
\label{sec:conclusion}

Merging subposteriors generated through parallel, independent MCMC simulations, to form the full posterior distribution is challenging. Currently, available methods either produce a Gaussian approximation to the posterior, or utilise nonparametric estimators which are difficult to tune and do not scale well to high-dimensional settings. In this paper, we have presented an alternative approach to this problem by directly modelling the log-density of the subposteriors. Using Gaussian-process priors, we were able to employ a fully Bayesian strategy towards approximating the full posterior, and unlike competing methods, we were able to account for the uncertainty in the approximation. 

Compared to the nonparametric methods, fitting the Gaussian-processes is straightforward using a mixture of marginalisation and maximum likelihood techniques for the hyperparameters. The main drawback of using Gaussian-process approximations is the computational cost. We have reduced the computational cost by, for each subposterior sample, thinning the Markov chain and removing duplicate MCMC samples prior to fitting the GP. We have shown that using only a small number of samples from the subposterior, we can accurately approximate the full posterior. Furthermore, the computationally intensive step of fitting the individual GPs to the subposteriors is automatically parallelised, as the subposteriors are independent by definition and the GPs are independent by design. While more computationally costly than some competing methods, it is important to note that the cost of fitting, and then sampling from the GP, is significantly cheaper than running an MCMC algorithm on the full data.

The results from Section \ref{sec:simulation-study} (and Appendices) show that in scenarios where both the subposterior and full posterior are approximately Gaussian, the consensus algorithm works well, and is computationally efficient to apply. In settings where either the subposteriors (mixture of Laplace distributions (Appendix \ref{sec:multi-modal-distr})) or full posterior (warped Gaussian (Section \ref{sec:warp-gauss-post}) and mixture model (Section \ref{sec:mixture-gaussians})) are non-Gaussian, our proposed Gaussian process approach is significantly superior to competing methods. This improved performance follows from using the log-subposterior densities to approximate the density of the full posterior, which other algorithms neglect to utilise.

The algorithms we propose scale well with the number of data points $n$, but fitting a GP when the dimension, $d$, of $\vartheta$ is high, can be computationally expensive as the number of input points required to produce an accurate approximation grows exponentially with $d$. We have explored the use of sparse GP approximations to reduce the computational burden and have shown that such approximations can be applied in this setting to produce faster algorithms with a similar level of accuracy as the standard GP. This is an ongoing area of research in the Gaussian process community and many alternative sparse GP approximations could be applied, potentially yielding improved results.

Finally, while not the focus of this work, we have numerically explored the effect of randomly partitioning the data. In scenarios where the dataset is heavily imbalanced (e.g. Bernoulli model from Section \ref{sec:rare-bern-events}), randomly partitioning the data can lead to non-overlapping subposteriors. This issue has not yet been addressed in the literature, and further work investigating ways to efficiently partition the data to ensure a more even distribution of the data across batches is ongoing. \\

\textbf{Acknowledgement}\\
We would like to thank Prof. Paul Fearnhead for helpful discussions. The first author gratefully acknowledges the support of the EPSRC funded  EP/H023151/1  STOR-i centre for doctoral training.

\bibliographystyle{apalike}
\bibliography{library}

\newpage
\appendix

\section{The HMC algorithm}
\label{app.HMC.alg}

\begin{algorithm}
   \caption{Hamiltonian Monte Carlo}
   \label{alg:hamiltonian}
\begin{algorithmic}
   \STATE {\bfseries Input:} Initial parameters $\vartheta^{(0)}$, step-size $\epsilon$ and trajectory length $L$.
   \FOR{$t=1$ {\bfseries to} $T$}
   \STATE Sample momentum $\varphi \sim \mathcal{N}(0,M)$
   \STATE Set $\vartheta_1 \leftarrow \vartheta^{t-1}$ and $\varphi_1 \leftarrow \varphi$
   \FOR{$l=1$ {\bfseries to} $L$}
   \STATE $\varphi_{l+\frac{1}{2}} \leftarrow \varphi_l + \frac{\epsilon}{2}\nabla_{\vartheta_l} \log \pi(\vartheta_l) $
   \STATE $\vartheta_{l+1} \leftarrow \vartheta_{l} + \epsilon M^{-1} \varphi_{l+\frac{1}{2}}$
   \STATE $\varphi_{l+1} \leftarrow \varphi_{l+\frac{1}{2}} + \frac{\epsilon}{2}\nabla_{\vartheta_{l+1}} \log \pi(\vartheta_{l+1}) $
   \ENDFOR
\STATE Set $\vartheta \leftarrow \vartheta_{L+1}$ and $\varphi \leftarrow \varphi_{L+1}$
\STATE With probability $\mbox{min}\left\{1,\frac{\exp \left(\log \pi(\vartheta) -\frac{1}{2}\varphi^\top M^{-1} \varphi \right)}{\exp \left(\log\pi(\vartheta^{t-1}) -\frac{1}{2}\varphi{^\top}^{(t-1)} M^{-1} \varphi^{(t-1)} \right)}\right\}$ \\ set $\vartheta^t \leftarrow \vartheta$ and $\varphi^t \leftarrow \varphi$ 
   \ENDFOR
  \STATE {\bfseries Output:} Samples $\{\vartheta^t\}_{t=1}^T$ from $\pi{(\vartheta)}$.
\end{algorithmic}
\end{algorithm}

\section{Additional results for the rare Bernoulli example}
\label{sec:addit-results-rare}

\begin{figure}[h]
  \centering
  \includegraphics[width=\textwidth,height=0.45\textheight]{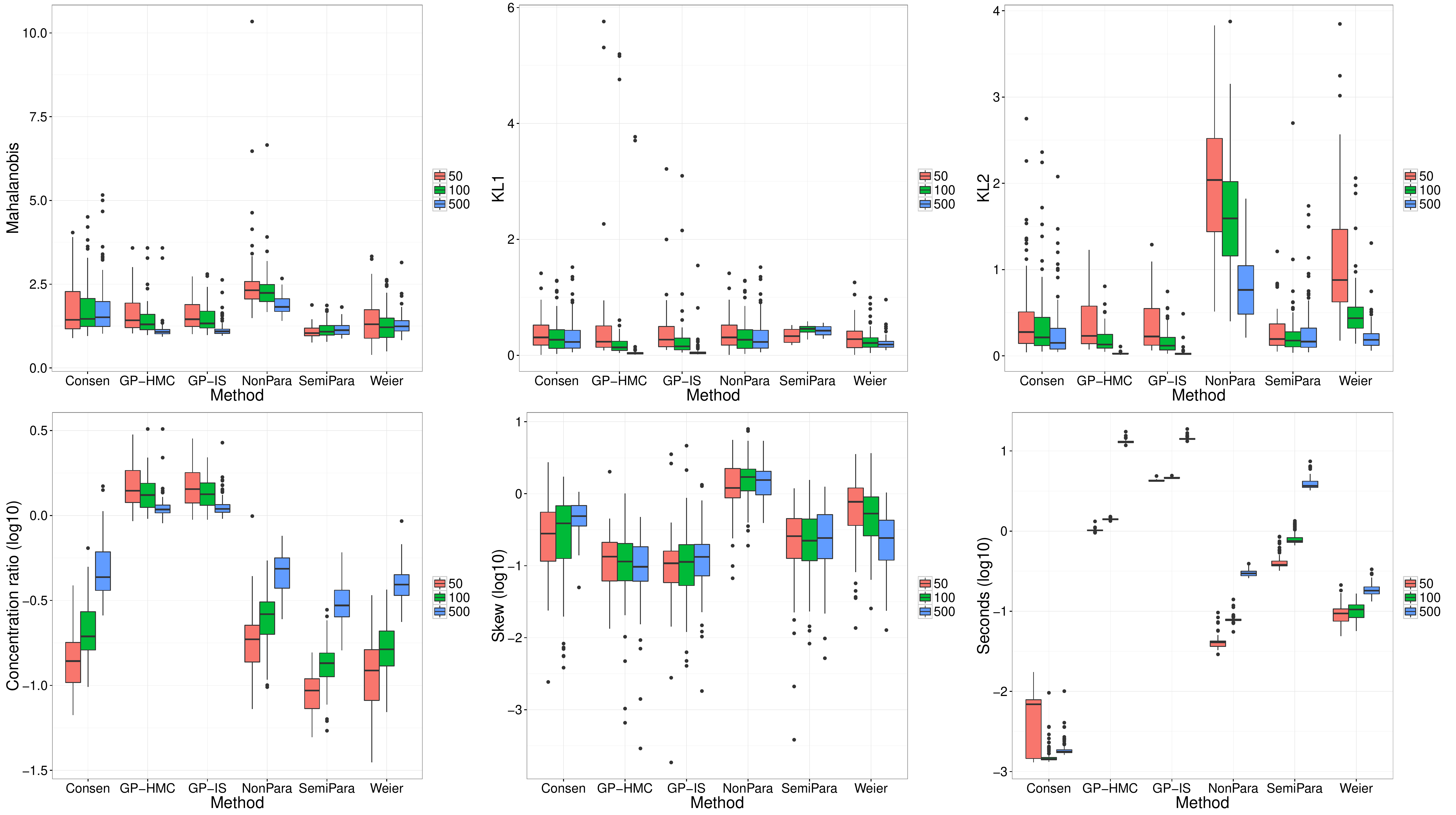}
\caption{}  
\label{fig:bernoulli_metrics}
\end{figure}

In Section \ref{sec:rare-bern-events} we considered the posterior distribution for rare Bernoulli random variables. We would expect that the accuracy of approximation would increase as the number of posterior samples, $J$, increases. In Figure \ref{fig:bernoulli_metrics} we consider the same discrepancy metrics from Section \ref{sec:simulation-study} and show how these are sensitive to posterior sample size, where we consider samples of size $J=(50,100,500)$. Notably, The GP approximations display considerable improvement as the sample size increases. This improvement, however, comes at an additional computational cost of fitting, and then sampling from, the GP. This cost, as measured in seconds, is significantly than the cost of the competing methods, but magnitudes of scale smaller than the cost of running the full MCMC algorithm on the whole dataset.

As noted in Section \ref{sec:rare-bern-events}, applying the distributed importance sampling step significantly improves the posterior approximation of all methods. Table \ref{tab:bernoulli-dis} gives the discrepancy metrics for each of the merge algorithms after a DIS step is performed.

\begin{table}
\centering
\begin{tabular}{l|ccccc}
\hline
Algorithm & $D_{Mah.}$ &$D_{KL}(\pi||\hat{\pi})$ & $D_{KL}(\hat{\pi}||\pi)$ & $\rho$ & $\eta$\\
\hline
Consensus      & $0.99$ & $\mathbf{0.66}$ & $\mathbf{0.46}$ & $1.02$ &  $\mathbf{0.06}$\\
Nonparametric  & $0.91$ & $0.89$ & $0.83$ &  $0.91$ & $0.14$\\
Semiparametric & $0.94$ & $0.79$ &$0.97$ & $0.82$ &  $0.16$\\
Weierstrass    & $1.05$ & $1.14$ & $1.16$ & $0.84$ & $0.28$\\
GP-HMC sampler &$\mathbf{0.85}$&  $0.80$& $0.76$ & $\mathbf{1.01}$ &$0.11$\\
\hline
\end{tabular}
\caption{Discrepancy metrics for the DIS step applied to each of the merge algorithms on a Bernoulli model
  with simulated rare event data. Kullback-Leibler results are
  reported as ($\times 10^{1}$) and $D_{Mah.}$ as ($\times
  10^5$). 
}
\label{tab:bernoulli-dis}
\end{table}

\section{Multimodal subposteriors}
\label{sec:multi-modal-distr}

\begin{figure}[h]
  \centering
  \includegraphics[width=0.45\textwidth,height=0.25\textheight]{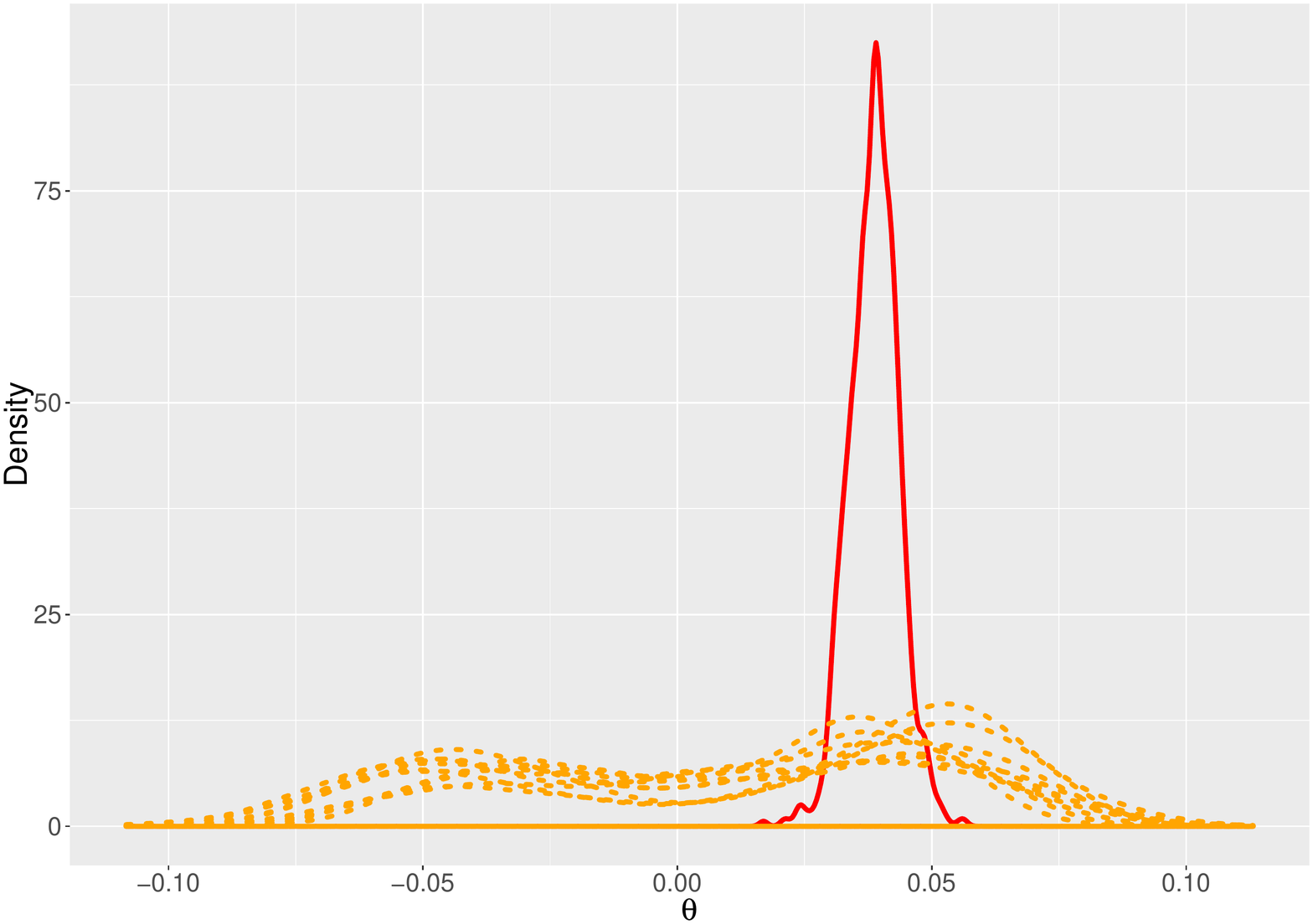}
  \includegraphics[width=0.45\textwidth,height=0.25\textheight]{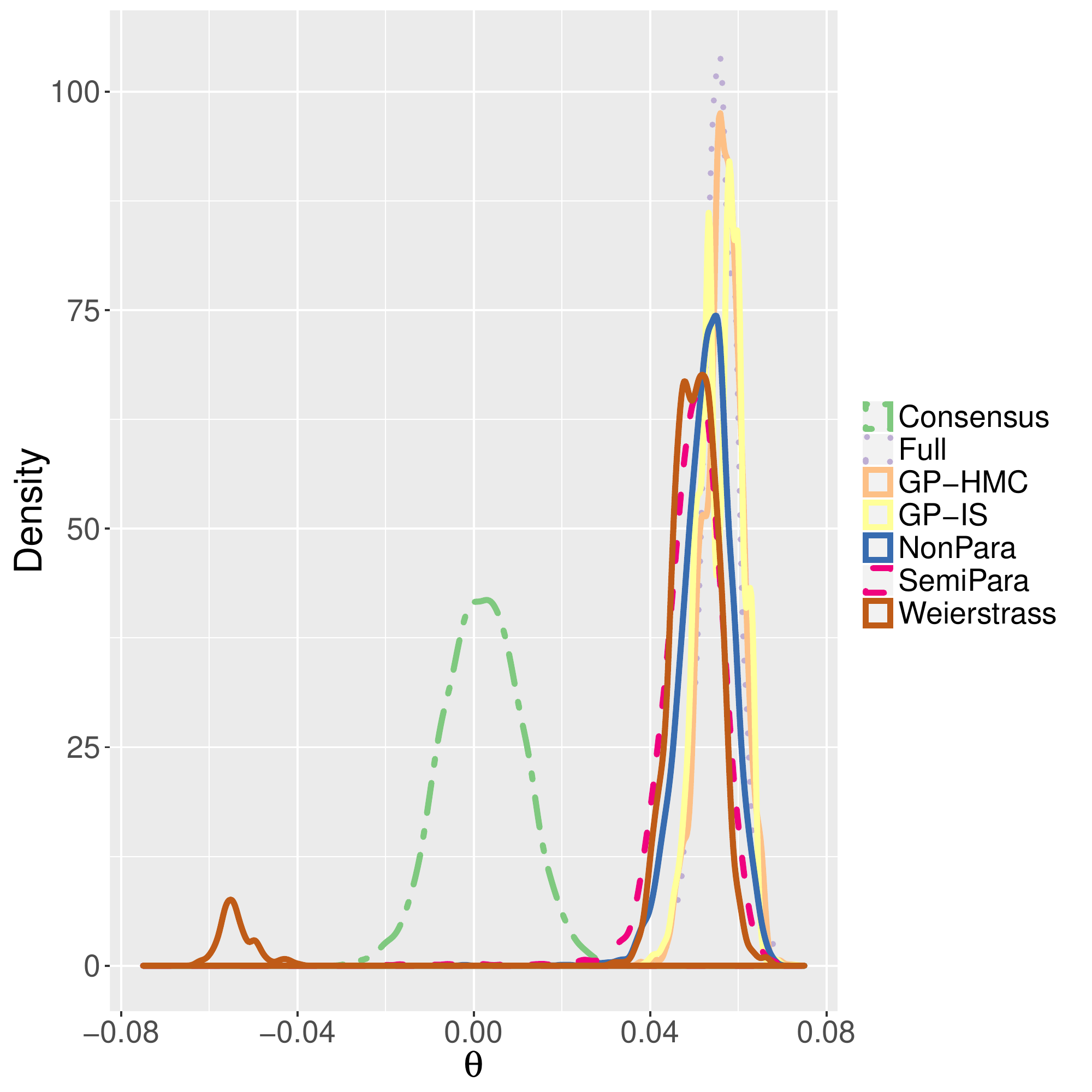}
  \caption{Left: The first ten subposteriors (orange dashed lines) plotted alongside the full posterior (red solid line). Right: Approximations of the full posterior given by each aggregation method.}
  \label{fig:multimodal}
\end{figure}

We create a concrete data scenario that could lead to a set of multimodal sub-posteriors similar to the artificial, perturbed multimodal sub-posteriors used in \cite{Wang2015}. The example is a toy representation of a general situation where one or more parameters of interest are poorly identified, but as the size of the dataset increases towards `large' $n$, the parameters start to become identifiable. The subposteriors are multimodal but the full posterior is unimodal (see Figure \ref{fig:multimodal}, left panel). 

We sample $n=1,000,000$ observations from a mixture of two Laplace distributions
\[
\frac{1}{2}\frac{1}{2\beta_1}\exp\left(-\frac{|y-\vartheta|}{\beta_1}\right) + \frac{1}{2}\frac{1}{2\beta_2}\exp\left(-\frac{|y+\vartheta|}{\beta_2}\right).
\]
If $\beta_1=\beta_2$ then the mixture components are non-identifiable,
however, by setting $\beta_1=1.01$ and $\beta_2=0.99$ the parameter of
interest, $\vartheta$ can be identified from a sufficiently large
dataset. For this experiment $\vartheta=0.05$ and the data are split
equally over $C=25$ processors. The scale parameters, $\beta_1$ and
$\beta_2$ are fixed at their true values and a $\mathcal{N}(0,1)$ prior is assumed
for $\vartheta$.

\begin{table}
\centering
\begin{tabular}{l|ccccc}
\hline
 Algorithm &$D_{Mah.}$& $D_{KL}(\pi||\hat{\pi})$ & $D_{KL}(\hat{\pi}||\pi)$& $\rho$  &$\eta$\\
\hline
Consensus      &$13.46$ ($5.62$)& $5.73$ ($5.27$) & $5.89$($5.16$) & $6.95$ ($2.41$) & $0.59$ $(3.13)$\\
Nonparametric  &$1.69$ ($1.66$)&  $0.26$ ($0.18$) & $0.36$ ($0.22$) & $1.92$ ($0.48$) & $0.56$ $(0.13)$\\
Semiparametric &$2.39$ ($1.79$)&  $0.77$ ($0.46$)& $0.73$ ($0.47$) & $1.98$ ($0.51$) & $1.78$ $(0.14)$\\
Weierstrass    &$6.52$ ($1.76$)&  $0.99$ ($0.75$)&  $0.91$ ($0.71$) &  $3.15$ ($0.40$) & $3.43$ $(0.27)$\\
GP-HMC sampler &$\mathbf{1.10}$ ($\mathbf{0.96}$)&  $\mathbf{0.15}$ ($\mathbf{0.13}$)& $\mathbf{0.19}$ ($\mathbf{0.16}$) & $1.09$ ($\mathbf{0.98}$) & $\mathbf{0.22}$ $(\mathbf{0.06})$\\
GP-IS sampler &$1.62$ &  $0.63$ & $0.42$  & $\mathbf{1.01}$ & $0.31$ \\
\hline
\end{tabular}
\caption{Accuracy of various aggregation methods on a mixture of
  Laplace distributions with a simulated dataset. $D_{Mah.}$ is
  reported as ($\times 10^5$). Results in brackets represent the DIS
  step.}
\label{tab:mulitmodal}
\end{table}

The left panel of Figure \ref{fig:multimodal} depicts the full posterior
and subposterior densities. The full posterior reveals,
approximately, 
the true value for $\vartheta$, whereas $\vartheta$ is poorly identified by
each of the subposteriors. The multimodality of the subposteriors
results in a poor posterior approximation from the consensus Monte
Carlo algorithm (right panel) as each subposterior is assumed to be
approximately Gaussian. On the other hand, most of the nonparametric
methods are able to capture the approximate shape of the posterior, but fail to correctly detect the posterior mode. Table
\ref{tab:mulitmodal} shows that the DIS step can slightly degrade the quality of the approximation if the proposal (e.g. semiparametric) under-represents the tail behaviour of the true posterior. As shown in Figure \ref{fig:multimodal} (right panel), the GP-HMC and GP-IS samplers produce good approximations to the full posterior and unlike the nonparametric methods, the GP approximations concentrate around the posterior mode (see $\rho$ in Table \ref{tab:mulitmodal}).

\section{Logistic regression model - simulated data}
\label{sec:logist-regr-model}

\begin{figure}[h]
  \centering
  \includegraphics[width=\textwidth,height=0.45\textheight]{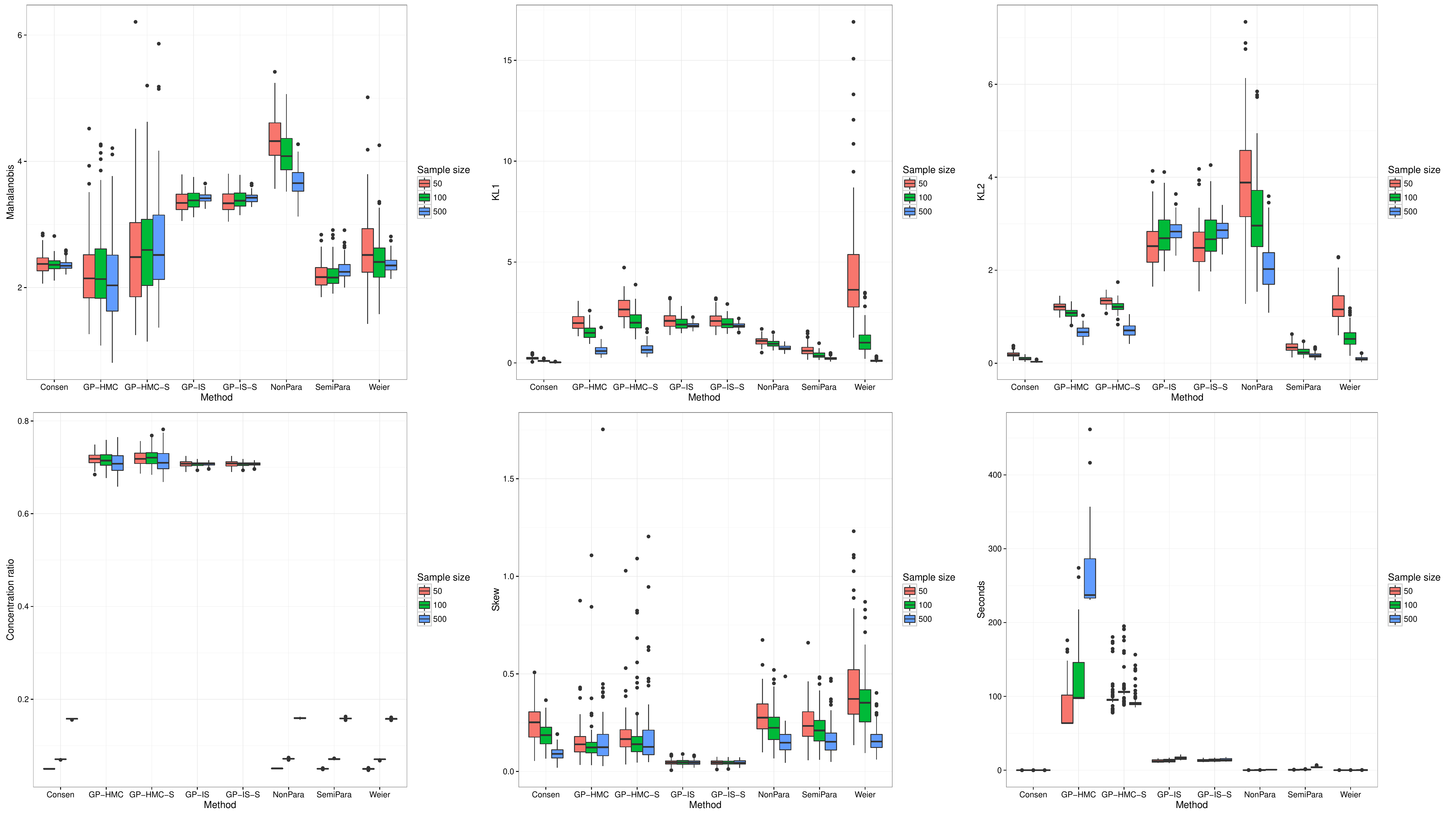}
\caption{Box plots for the discrepancy metrics corresponding to the simulated data example for the logistic regression model from Section \ref{sec:logistic-regression}. }  
\label{fig:logistic_simulated}
\end{figure}

In Figure \ref{fig:logistic_simulated} we compare the various divide-and-conquer MCMC strategies with varying sample size. Each algorithm performs reasonably well, as reported in Section \ref{sec:logistic-regression}. All of the methods show an improvement when the sample size increases, however, the Gaussian process type algorithms less so than the consensus and nonparametric type algorithms. On balance, this suggests that, for the additional computational cost of fitting the GP, increasing the sample size may not provide additional posterior accuracy when accounting for computational cost.

\begin{figure}[h]
  \centering
  \includegraphics[width=\textwidth,height=0.35\textheight]{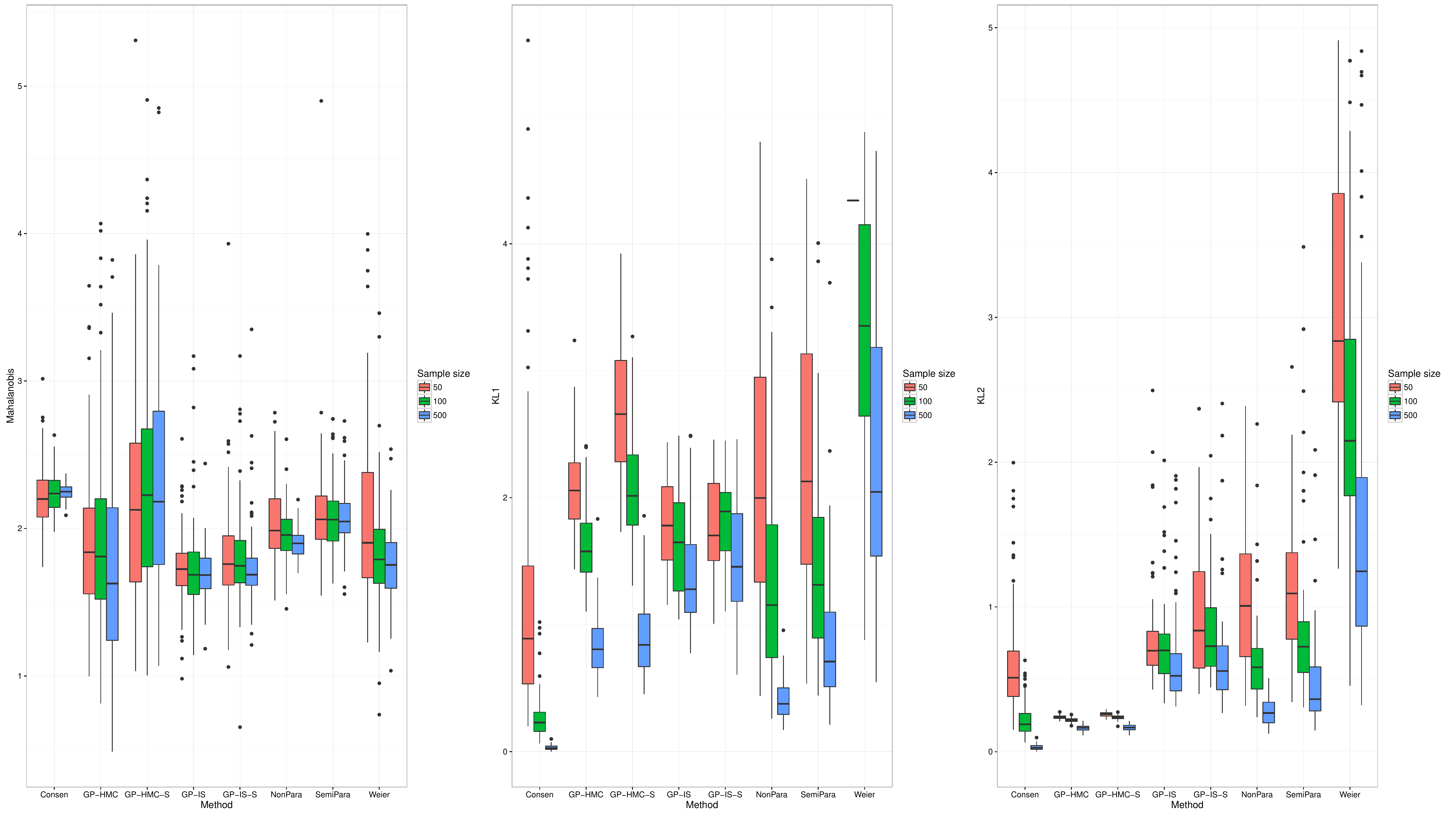}
\caption{Box plots for the discrepancy metrics corresponding to the simulated data example for the logistic regression model from Section \ref{sec:logistic-regression} after distributed importance sampling is applied.}  
\label{fig:logistic_simulated_dis}
\end{figure}

Applying the distributed importance sampling step acts as a correction for all of the divide-and-conquer algorithms (Figure \ref{fig:logistic_simulated_dis}). This step offers particular benefit for the nonparametric based algorithms, which typically scale poorly with dimension.

\section{Logistic regression model - Hepmass data}
\label{sec:logist-regr-model-1}

\begin{figure}[h]
  \centering
  \includegraphics[width=\textwidth,height=0.45\textheight]{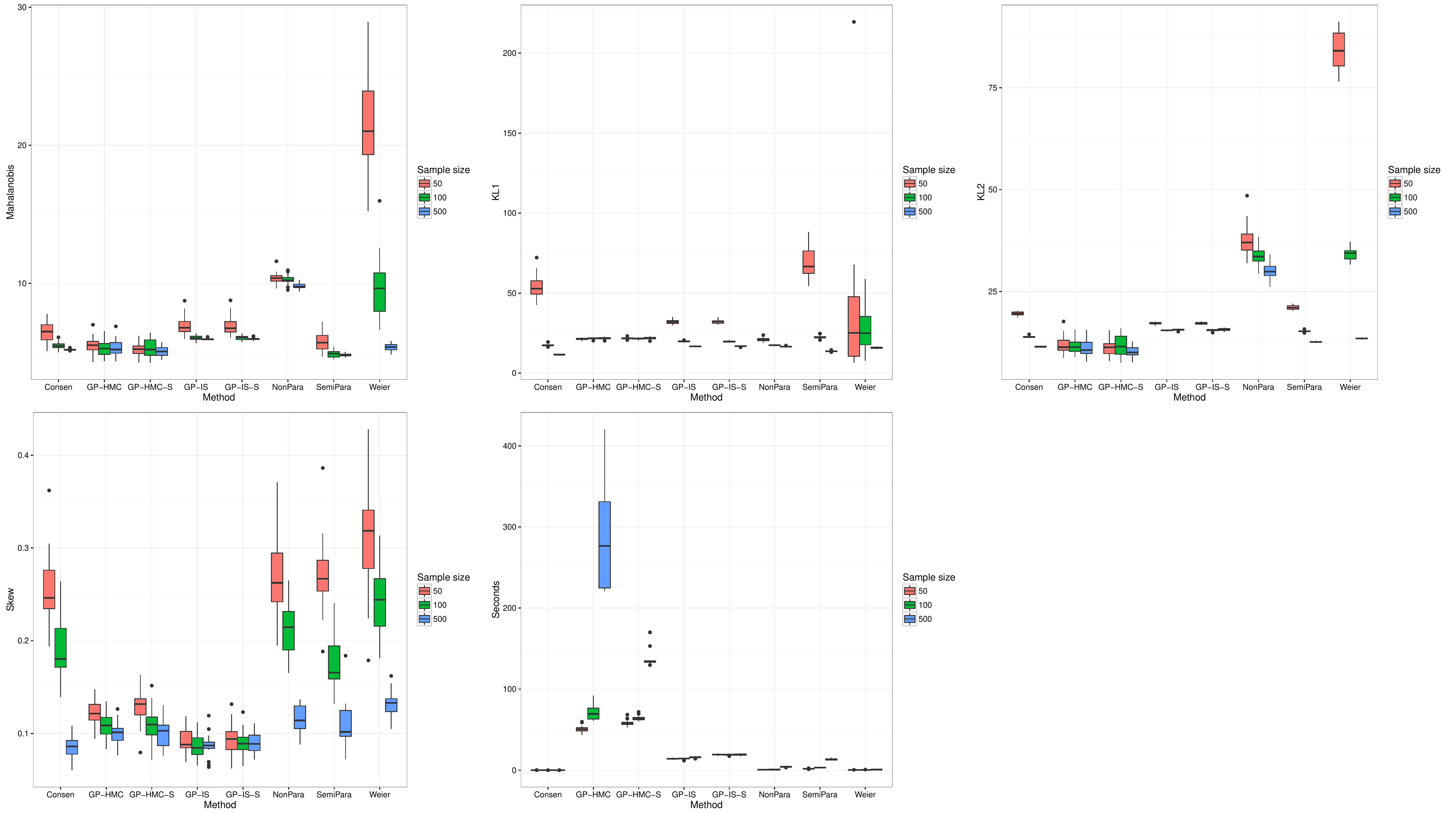}
\caption{Box plots for the discrepancy metrics corresponding to the hepmass data example for the logistic regression model from Section \ref{sec:logistic-regression}. }  
\label{fig:logistic_hepmass}
\end{figure}

Figure \ref{fig:logistic_hepmass} provides box plots for each of the discrepancy metrics and for each divide-and-conquer algorithm. Increasing the sample size in this example leads to a more significant improvement in the discrepancy metrics compared to the simulated data example (Figure \ref{fig:logistic_simulated}). This difference is related to the dimensionality of the posterior distribution (28 parameters in the hepmass example, 5 in the simulated example). The improvement is most notable for the consensus algorithm, for small sample sizes the Gaussian assumption underlying the consensus algorithm is not valid. As the sample size increases, the Gaussian assumption becomes more realistic.

\end{document}